\pdfoutput=1
\documentclass[10pt,conference,compsocconf]{IEEEtran}
\usepackage[linesnumbered]{algorithm2e}
\usepackage{amssymb}
\usepackage{mathtools}
\usepackage{graphicx}
\usepackage{gensymb}
\usepackage{xcolor}
\usepackage[utf8]{inputenc}
\usepackage{xspace}
\usepackage{booktabs}
\usepackage{cleveref}
\usepackage{url}
\usepackage{times}
\usepackage{subcaption}

\usepackage{flushend}

\newcommand{\eg}{e.g.,\xspace}

\newcommand{\etc}{etc.\@\xspace}
\newcommand{\ie}{i.e.,\xspace}

\newcommand{\unit}[1]{\ensuremath{\, \mathrm{#1}}}
\newcommand{\atomic}[0]{atomic section\xspace}
\newcommand{\suitename}{ESCAPE\xspace}

\graphicspath{{./figures/}}

\begin{document}
\title{Enabling Secure Location Verification for Urban Road Applications}
\title{Security of GPS/INS based On-road Location Tracking Systems}
\author{\IEEEauthorblockN{
	Sashank Narain\IEEEauthorrefmark{1},
	Aanjhan Ranganathan\IEEEauthorrefmark{2} and
	Guevara Noubir\IEEEauthorrefmark{3}}

\IEEEauthorblockA
	{College of Computer and Information Science\\
	  Northeastern University, Boston, MA, USA\\
	Email: 
\IEEEauthorrefmark{1}sashank@ccs.neu.edu,
\IEEEauthorrefmark{2}aanjhan@northeastern.edu,
\IEEEauthorrefmark{3}noubir@ccs.neu.edu}}
\maketitle
\begin{abstract}
Location information is critical to a wide-variety of navigation and
tracking applications. Today, GPS is the de-facto outdoor localization
system but has been shown to be vulnerable to signal spoofing
attacks. Inertial Navigation Systems (INS) are emerging as a popular
complementary system, especially in road transportation systems as
they enable improved navigation and tracking as well as offer
resilience to wireless signals spoofing, and jamming attacks. In this
paper, we evaluate the security guarantees of INS-aided GPS tracking
and navigation for road transportation systems. We consider an
adversary required to travel from a source location to a destination,
and monitored by a INS-aided GPS system. The goal of the adversary is
to travel to alternate locations without being detected. We developed
and evaluated algorithms that achieve such goal, providing the
adversary significant latitude. Our algorithms build a graph model for
a given road network and enable us to derive potential destinations an
attacker can reach without raising alarms even with the INS-aided GPS
tracking and navigation system. The algorithms render the gyroscope
and accelerometer sensors useless as they generate road trajectories
indistinguishable from plausible paths (both in terms of turn angles
and roads curvature). We also designed, built, and demonstrated that
the magnetometer can be actively spoofed using a combination of
carefully controlled coils.  We implemented and evaluated the impact
of the attack using both real-world and simulated driving traces in
more than 10 cities located around the world. Our evaluations show
that it is possible for an attacker to reach destinations that are as
far as 30 km away from the true destination without being detected. We
also show that it is possible for the adversary to reach almost
60--80\% of possible points within the target region in some
cities. Such results are only a lower-bound, as an adversary can
adjust our parameters to spend more resources (e.g., time) on the
\textit{target} source/destination than we did for our performance
evaluations of thousands of paths.  We propose countermeasures which
can severely limit an attackers ability without the need for any
hardware modifications. For instance, our system can be used as the
foundation for countering such attacks, both detecting and
recommending paths that are difficult to spoof.

\end{abstract}

\section{Introduction}
The ability to track one's location is important to a wide variety of safety- and security-critical applications. For example, logistics and supply chain management companies~\cite{coffee2003vehicle,novik2002system,verizon-fleet-connect} that handle high-value commodities (e.g., currency notes) continuously monitor the locations of every vehicle in their fleet carrying valuables to ensure their secure transportation to the intended destination. Emergency support services such as medical and law enforcement rely on location information to track their personnel, optimize response times and to even activate traffic signal lights appropriately. Law enforcement officials use ankle bracelets~\cite{ma-electronicmonitoring,geosatis} to monitor the location of defendants or parole and notify them if the offender strays outside an allowed area. Ride-hailing applications such as Uber and Lyft use location information for tracking, billing, and assigning drivers to trips. Furthermore, the locations of public transport~\cite{masstransit-usa,mintsis2004applications,civitas_eu} are continuously monitored to ensure smooth and timely operation of services. With the advent of autonomous vehicles and transport systems, the dependence on location information is only bound to increase. The majority of above applications rely on Global Positioning Systems (GPS)~\cite{misra2006global} which is the de facto outdoor localization system in use today. It is estimated that more than 8 billion GNSS\footnote{Global Navigation Satellite Systems (GNSS) is an umbrella term for satellite based localization systems such as GPS, Galileo, Glonass etc.} devices~\cite{gsa2017market} will be in use by the year 2020. 

However, it has been widely demonstrated that GPS is vulnerable to signal spoofing attacks. One of the main reasons is the lack of any form of signal authentication. It is today possible to change the course of a ship~\cite{yacht_spoofing}, force a drone to land in an hostile area~\cite{humphreys2012statement} or fake the current location in a road navigation system~\cite{zeng2017practical} by simply spoofing GPS signals. The increasing availability of low-cost radio hardware platforms make it feasible to execute such attacks with less than few hundred dollars worth of hardware equipment. There has been several evidences of jamming and spoofing reported in the media. For example,~\cite{gizmodo_gpsjamming} quotes ``Because the toll-taking for commercial trucks relies on GPS tracking, they can avoid paying through jamming. If a \$45 device made your daily commute free, you too might be tempted to commit a federal crime.'' Another report~\cite{cbs_gpsjamming} mentions ``Gary Bojczak admitted buying an illegal GPS jammer to thwart the tracking device in his company vehicle''. Several countermeasures have been proposed in the recent years either to detect or to mitigate signal spoofing attacks. Cryptographic mitigation techniques~\cite{humphreys2013detection,kuhn2005asymmetric,lo2010authenticating,wesson2012practical} (e.g., military GPS systems where the spreading codes are secret) require changes to the satellite infrastructure. Furthermore their use requires distribution and management of shared secrets, which makes them impractical for majority of applications. Non-cryptographic countermeasures~\cite{Akos2012,ranganathan2016spree,psiaki2013gnss,Wesson2011,Warner2003,broumandan2012gnss,Jafarnia-Jahromi2012} rely on identifying anomalies in the physical characteristics of the received GPS signal. These techniques are either unreliable (e.g., large number of false alarms), effective only against naive attackers or require modifications to the GPS receiver itself. Alternate localization technologies using WiFi or cellular networks~\cite{zandbergen2009accuracy,tippenhauer2009attacks} lack the accuracy and coverage required for the above mentioned applications. Moreover, they consume significant amount of power and are susceptible to external signal and environmental interference. 

Inertial navigation i.e., the use of sensors such as accelerometer, gyroscope and compass to navigate during temporary GPS outages have been around for decades, specifically in aircrafts, spacecrafts and military vehicles~\cite{titterton2004strapdown,farrell1999global,wendel2006integrated}. The advancements in sensor manufacturing technologies have resulted in widespread integration of these sensors into many commonly used devices such as smart phones, tablets, fitness trackers and other wearables. Many vehicle tracking and automotive navigation systems have integrated GPS with inertial measurement units to improve localization and tracking of individual vehicles~\cite{kvh_imu,vectorNAV_imu,honeywell_IMU,navtech_gpsimu}. Inertial sensors are key to the balancing and navigation technologies present in modern segways. Low-cost inertial sensors have also proliferated into the consumer drone industry today.
One of the key advantages of inertial navigation is its robustness and resilience to any form of wireless signal spoofing and jamming attacks as there is no need for the sensors to communicate or receive information from any external entity such as satellites or other terrestrial transponders. This makes them very attractive for use in security- and safety-critical localization and tracking applications where GPS (or any wireless) spoofing and jamming attacks are a concern. The main drawback of inertial navigation units is the accumulating error of the sensor measurements. These accumulated sensor measurement errors affect the estimated position and velocity over a longer duration of time and hence limit the maximum period an inertial unit can act independently. This affects aerial and maritime navigation capabilities significantly as the tracked vehicle has all the six degrees of freedom to move. However, in the context of road navigation, the vehicle is limited by the road network and can only navigate within the constraints of these existing roadways. These inherent constraints imposed by the road networks have made low-cost inertial sensors very valuable for quick attack detection and immediate tracking of cheating entities~\cite{khanafseh2014gps,white1998detection,lee2015gps,dehnie2014methods,ebner1997integrated,serrano2011receiver}.

In this work, we evaluate the security guarantees of GPS/INS based on-road location tracking systems. Specifically, we address the following research questions: Given a geographic area's road network and assuming that both GPS and inertial sensor data are continuously monitored for tracking an entity's location, is it possible for an attacker to fake its navigation path or final destination? If yes, what are the attacker's constraints and possibilities? Can we exploit the physical motion constraints that exist in an urban road network and design a secure navigation algorithm that generates travel routes that are hard to spoof? For example, can a driver of a vehicle carrying high-value commodities (e.g., currency notes) spoof his assigned route and deviate without being detected by the monitoring center? Can a parole with GPS/INS ankle monitor spoof his location and travel routes without causing any discrepancies in the estimates computed by both GPS and inertial sensors?

Specifically, we make the following contributions in this
paper. First, we demonstrate that GPS/INS based on-road location
tracking and navigation has severe limitations. We develop algorithms
and a system that show it is indeed possible for an attacker to hijack
vehicles far away from the intended destination or take an alternate
route without triggering any alarms even though the GPS location as
well as inertial sensors are continuously monitored. We leverage the
regular patterns that exist in urban road networks and create a suite
of algorithms which we refer to as \suitename that automatically
suggests potential routes to spoof given a start point $s$, and end
point $d$. The paths are generated to be highly plausible to travel
from $s$ to $d$, yet easy to spoof at the INS sensors levels. Spoofing
means that the adversary will travel on an alternate path
indistinguishable from the spoofed path.  Our \suitename suite of
algorithms provides possible escape routes an attacker can take
without being detected while spoofing. It incorporates intersections
turn angles, roads curvatures, and magnetometer bearing. We evaluated
our attack's feasibility and impact in 10 major cities across the
globe and the results show that an attacker can potentially take the
vehicle as far as 30 km before the monitoring system can detect a
potential attack. Note that even after detection, the tracking system
has no knowledge of the true location. To the best of our knowledge
this is the first demonstration of the security vulnerabilities that
exist in GPS/INS based location verification and tracking systems. Our
attack affects several services and applications with effective
monetary value running into several millions of dollars. Our attacks
essentially renders the gyroscope and accelerometer useless by
generating paths acceptable to the monitoring system, but have a
signature indistinguishable from the trajectory effectively traveled
by the adversary. For the magnetometer, a sensor that can play a
critical role in detecting the incongruence of the claimed trajectory
with the measured heading, we built and demonstrated the effectiveness
of a magnetometer-spoofing device that physically generate a magnetic
field compatible with the spoofed trajectory. Finally, based on the
observations, we turn around our \suitename suite of attack algorithms
to build a countermeasure that the tracking services can run to
mitigate such spoofing attacks. Specifically, we modified \suitename
to output secure navigation routes that can be assigned given a start
and end points that severely limits the attacker's possibilities.




\section{Background}

\subsection{Overview of GPS}
GPS is today the de-facto outdoor localization system used. GPS is a satellite-based global navigation system that consists of more than 24 satellites orbiting the earth at more than 20,000~km above the ground. Each satellite is equipped with high-precision atomic clocks and hence the timing information available from the satellites are in near-perfect synchronization. Each satellite transmits messages referred to as the \textit{navigation messages} that are spread using pseudorandom codes unique to that satellite. The GPS receiver on the ground receives these navigation messages and estimates their time of arrival.  Based on the time of transmission contained within the navigation message and its time of arrival, the receiver computes its distance to each of the visible satellites. Once the receiver acquires the navigation messages from at least four satellites, the GPS receiver estimates its own location and precise time using the standard technique of multilateration.

\subsection{GPS Spoofing Attacks}
Civilian GPS is easily vulnerable to signal spoofing attacks due to the lack of any signal authentication and the publicly known spreading codes for each satellite, modulation schemes,	 and data structure. A GPS signal spoofing attack is a physical-layer attack in which an attacker transmits specially crafted radio signals that are identical to authentic satellite signals. In a signal spoofing attack, the objective of an attacker may be to force a target receiver to (i) compute a false geographic location, (ii) compute a false time or (iii) disrupt the receiver by transmitting unexpected data. Due to the low power of the legitimate satellite signal at the receiver, the attacker's spoofing signals can trivially overshadow the authentic signals. During a spoofing attack, the GPS receiver locks onto (acquires and tracks) the stronger signal \ie the attacker's signals, ignoring the legitimate satellite signals. This results in the receiver computing a false position, velocity and time based on the spoofing signals. Today, with the increasing availability of low-cost radio hardware platforms~\cite{ettus_research,lowcost_gps} and open source GPS signal generation software~\cite{gps-sdr-sim}, it is feasible to execute GPS spoofing attacks with less than \$100 of hardware equipment. GPS signal generators can be programmed to transmit radio frequency signals corresponding to either a static position (e.g., latitude, longitude and elevation) or simulate entire motion trajectory. For example, an attacker can spoof the navigation route of a vehicle carrying high-value items and hijack it to any arbitrary location without rising any alarms. The operators of ride hailing services can fake the route taken for a trip. Furthermore, GPS spoofing attacks can delay or even prevent emergency support services from reaching the intended destinations. Given the implications of GPS spoofing attacks on road navigation and tracking applications, it is essential to ensure resilience against these modern day cyber-physical attacks.

\begin{figure}[t]
	\centering
	\begin{subfigure}{.45\linewidth}
        \centering
		\includegraphics[width=\textwidth]{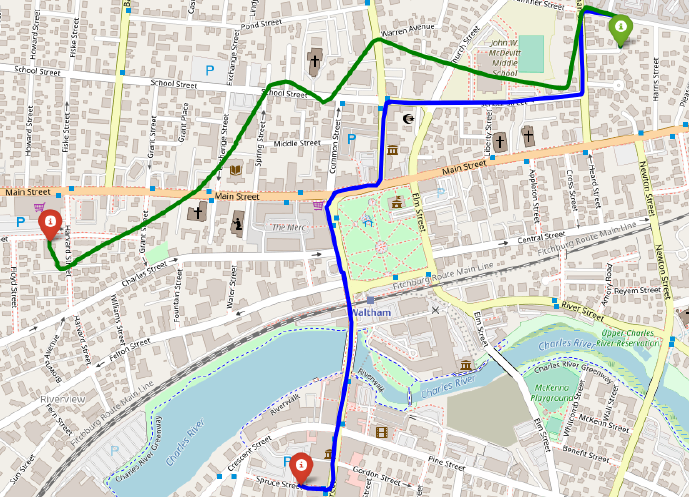}
		\caption{Path drift}
    \end{subfigure}%
    \begin{subfigure}{.55\linewidth}
        \centering
		\includegraphics[width=\textwidth]{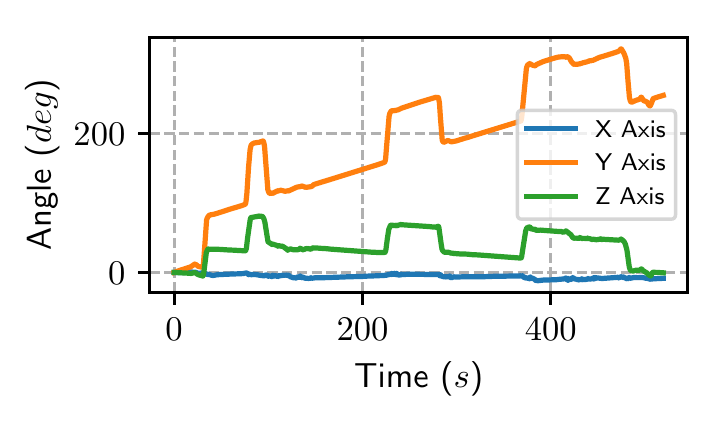}
		\caption{Gyroscope drift}
    \end{subfigure}
	\caption{The constraints imposed by the road networks lead to better accuracy in tracking road applications. The blue path is the actual and estimated route taken by a vehicle and tracked using low-cost inertial sensors. The green path is the estimated trajectory in case of aerial navigation.}
	\label{fig:path_drift}
\end{figure}

\subsection{Inertial Sensors Aided Navigation and Tracking}
The need to operate effectively in scenarios where GPS is inaccessible, unreliable or potentially jammed or spoofed by adversaries has led to the increased interest in building complementary navigation solutions and spoofing detection techniques. Several countermeasures and alternative localization techniques have been proposed. Of them, inertial sensors are emerging as a popular choice for two main reasons. First, inertial measurements are not affected by wireless signal jamming and are therefore resilient to denial of service attacks. Second, their widespread availability in majority of modern smartphones makes them easy to deploy and integrate into existing navigation and tracking infrastructure without the need for any hardware or software modifications to the GPS receiver. 

Inertial navigation is the process of integrating the readings of select sensors such as accelerometers, gyroscopes, and magnetometer into a complete three-dimensional position, velocity, and orientation solution. Inertial navigation systems are classified as \emph{dead-reckoning}, since the estimation process is iterative and uses prior information \ie calculating from some previously known navigation solution. Accelerometers measure both gravitational and non-gravitational acceleration along each of the three axes. The gyroscopes measure the rate at which an object is rotating, and are used to compute the attitude and heading of the object. The gyroscope measurements aid the accelerometer in figuring out the orientation of the object. Typically, sets of three accelerometers and three gyroscopes, both orthogonally aligned, are usually combined into a single inertial measurement unit (IMU), which commonly contains additional analog and digital circuitry, including conversion and calibration components. As the name implies, the magnetometer measures the magnetic fields and thus determine the cardinal direction to which the object is pointing.

One of the main drawbacks of low-cost inertial sensors (e.g., MEMS~\cite{rebeiz2004rf}) is that the process of dead reckoning in general, results in a build-up of errors over the course of the measurement. Since the position, velocity, and attitude updates are products of single or double integration of raw inertial sensor readings, the errors propagate and affect the final position, velocity and attitude estimates. For example, due to the single integration performed on angular rate measurements, a constant gyroscope bias will produce a linearly growing angular error, the gyro noise will produce a `random walk' growing with the square root of time. The double integration required to transform the accelerometer output to position produces a quadratically growing position error and a second-order `random walk', for a constant accelerometer bias and white noise respectively. In numerical terms, a $25 \unit{\mu m^2 s^{-1}}$ accelerometer bias ($\approx 245 \unit{\mu g}$) of a navigation grade sensor would produce a $1.59\unit{km}$ position error in one hour. The aggravation of sensor errors becomes critical to aviation and maritime applications as the vehicle have more degrees of freedom to move. However, on road, the vehicles are limited by the available road networks and are therefore severely constrained in their possible trajectories. \Cref{fig:path_drift} illustrates how the bias errors affect the final position estimates in a road navigation scenario (with motion constraints) and aerial (without any motion constraints). These constraints imposed inherently by the road networks has led to the emergence of using inertial sensors to complement GPS navigation and tracking solutions. Moreover, the inertial sensors are largely immune to jamming which makes them invaluable to the safety and security-critical applications described previously.

In this paper, we focus on the security of such on-road systems that rely on both GPS and inertial sensor measurements for navigation and tracking. We begin with demonstrating how an attacker can fake his navigation route even if both the GPS and the inertial sensors are continuously monitored in the next section.
%
%
%

\section{Spoofing INS-aided Localization Systems}
In this section, we demonstrate spoofing attacks on road navigation
and tracking applications that rely on both GPS and the inertial
sensors for the localization. To the best of our knowledge, this is
the first demonstration of spoofing attacks on GPS/INS localization
systems. First, we describe the system and attacker model. Then, we
give a high-level overview of the proposed spoofing attack algorithms
and define relevant terminologies. Finally, we describe in detail the
working of our attack algorithms.

\subsection{System and Attacker Model}
In this work, we focus on localization and tracking systems that rely
on both GPS and INS measurements to navigate and track entities. As
described previously, such GPS/INS systems are gaining popularity in
road navigation and tracking applications due to the improved
accuracy, availability and resilience to signal jamming/spoofing
attacks. Our attack is independent of how the GPS/INS system is
deployed \ie it can either be an app on a trusted smartphone or a
specialized tracking device (e.g., ankle monitors) installed on the
entity of interest. The main objective of the monitoring system is to
keep track of the location and navigation routes of the entities. We
assume an attacker capable of generating and transmitting fake GPS
signals corresponding to any location or navigation route of his
choice using tools such as GPS-SDR-SIM~\cite{gps-sdr-sim}. The goal of
the attacker is to spoof his location and navigation trajectory
without being detected. For example, the attacker can try to deviate
from an assigned navigation route and reach as far away as possible
from the intended destination before an anomaly is detected and an
alarm raised. At that moment, the adversary's location remains
undetermined. Alternately, the attacker starts and ends at the
intended locations, however using a different route than the one being
reported to the monitoring station. We assume that the attacker has
full physical access to the entity being tracked and is aware of the
GPS/INS system deployed for monitoring. However, we assume that the
tracking device itself is tamper-proof. For example, the attacker can
be a driver of a cargo company (or a hijacker) who has full access to
the vehicle. He regularly drives this vehicle to transport high-value
goods, and is aware of the GPS and INS based tracking system employed
by the company. However, he cannot modify the software on the
smartphone or physically tamper the tracking device.




\subsection{Overview of the Attack}
The primary objective of the attacker is to fake the reported
navigation route without raising suspicion of any mischief. Note that
simply spoofing GPS signals is not sufficient as the INS measurements
will indicate discrepancies between the reported GPS location and the
inertial estimates. In order to successfully execute the attack, it is
now necessary for the attacker to identify and spoof navigation paths
that have similar distances, road curvature, and turn angles to
minimize the discrepancies between the INS and GPS estimates. Our
system, which we refer to as \suitename, exploits the regular patterns
that exist in many cities' road networks and identifies navigation
paths that are similar to the route that is reported to the monitoring
center. As a result, the inconsistencies between the INS and GPS
estimates are negligible and the attack is successfully executed.

\begin{figure}[t]
  \centering
  \includegraphics[width=0.52\columnwidth]{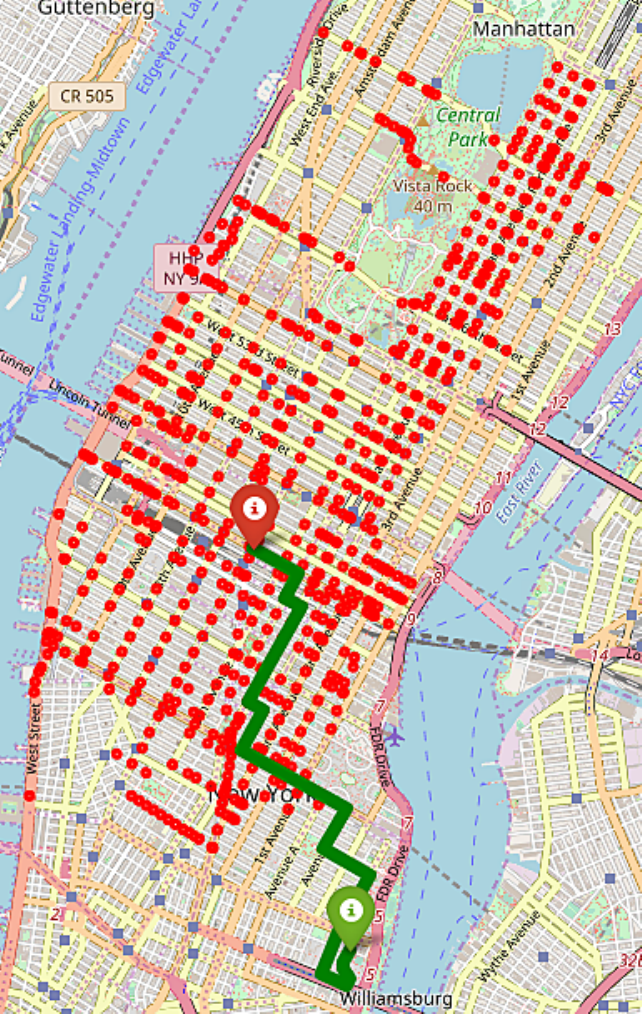}
  \caption{An example of a spoofed path in Manhattan and the escape
    destinations generated for that single spoofed path. Our
    algorithms generate 100 spoofed paths for a given
    source/destination locations, allowing an adversary to
    undetectably reach an even larger set of location.}
  \label{fig:spoofed_destinations}
\end{figure}

The attack begins with the attacker providing the start and end points
of the assigned trip to \suitename. Then, \suitename computes two sets
of paths: (i) \emph{spoofed paths} and (ii) \emph{escape paths}. The
\textit{spoofed paths} are a set of paths that exist between the input
start and end points of the trip. These are the paths that the
attacker will generate fake GPS signals and spoof the receiver to
report to the monitoring center. These should be plausible paths for
the source and destination locations, and not raise suspicion. For
every spoofed path, \suitename computes a set of \textit{escape paths}
which the attacker can use to deviate from the intended course while
executing the spoofing attack. In other words, a \emph{spoofed path}
is the route that is reported to the monitoring center and the
\textit{escape path} is the true route taken by the attacker to reach
an alternate destination. The attacker then picks an escape path that
enables him to reach his intended location. The intended location can
either be a point far away from the assigned destination (to buy the
adversary some time) or just a diversion before reaching the assigned
destination. The selected escape path corresponds to a spoofed path
which the attacker can use to generate spoofing
signals. \Cref{fig:spoofed_destinations} illustrates an example of a
spoofed path generated between two end points in Manhattan (green line
from green marker to red marker) and the destinations of the escape
paths (red points) generated for this particular spoofed
path. Finally, the attack is executed by spoofing the tracking device
to report the \emph{spoofed path} while the attacker actually drives
the \emph{escape path}. In the next section, we present the inner
working of our \suitename attack system.

\subsection{Internals of \suitename}

\suitename consists of three main building blocks: (i) graph
constructor, ii) spoofed paths generator and (iii) escape paths
generator. The graph constructor generates directed graphs based on
the road network present in the geographic area of interest. Our
attack does not enforce any limits on the geographic area. As the name
suggests, the spoofed and escape paths generator blocks are
responsible for computing and identifying spoofed and escape paths for
the attacker.

\subsubsection{Graph Constructor}
\label{sec:graph}

\begin{figure}[t]
	\centering
	\begin{subfigure}{\linewidth}
        \centering
		\includegraphics[width=.65\textwidth]{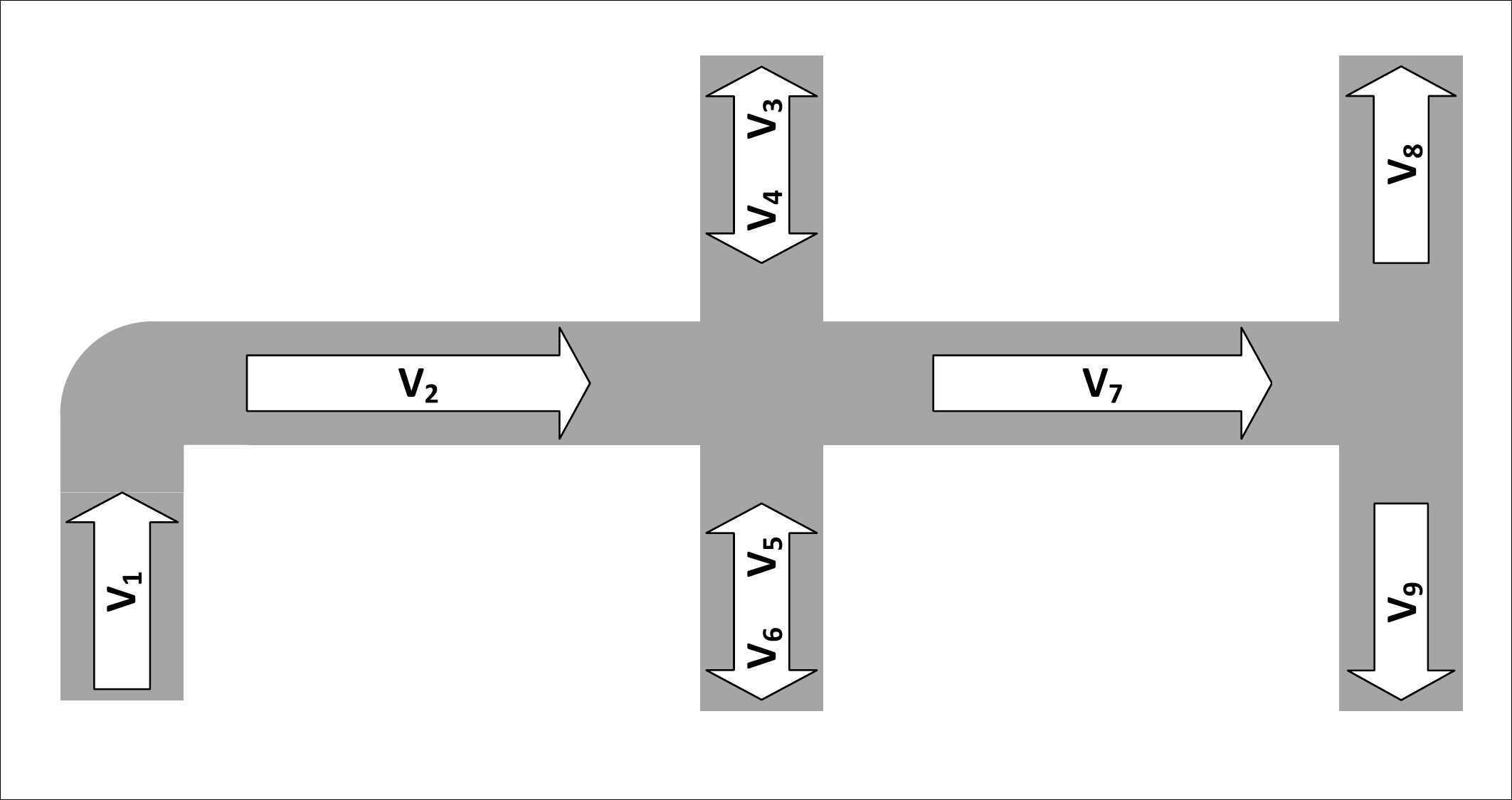}
		\caption{\label{fig:roadnetwork} Example Road Network}
    \end{subfigure}
    \begin{subfigure}{\linewidth}
        \centering
		\includegraphics[width=.58\textwidth]{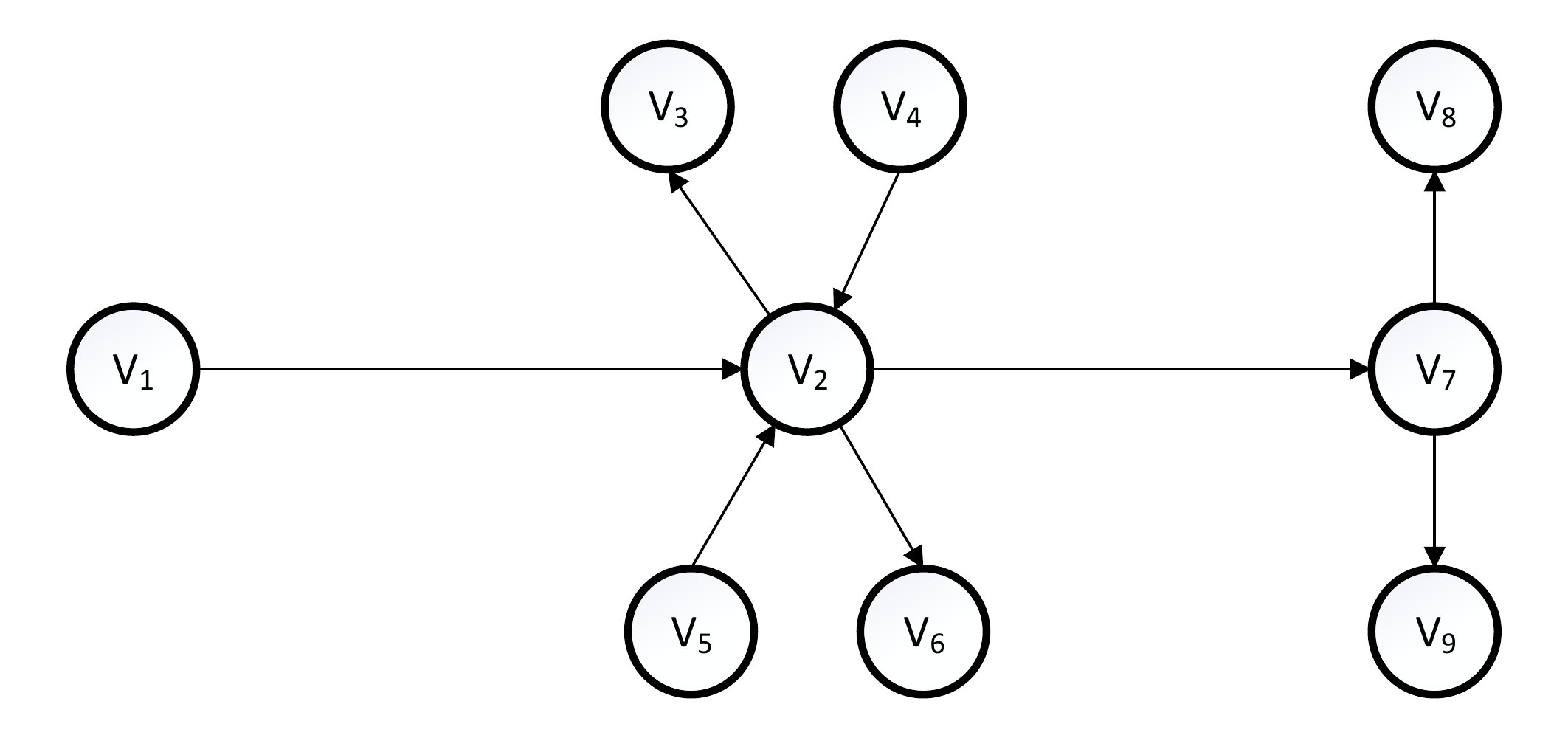}
		\caption{\label{fig:graph} The Graph Representation}
    \end{subfigure}
	\caption{Example of a road network and its corresponding graph representation.}
	\label{fig:graph}
\end{figure}

The paths for a geographic area $\mathcal{G}$ are generated from a
directed graph $G_\mathcal{G} = (V, E)$. We chose
OpenStreetMap~\cite{osm} as the map provider because it contains
accurate road information for all major cities of the world along with
various meta-data such as types of roads and buildings. Each
geographic area can be represented as
$\mathcal{G}=(\mathcal{A},\mathcal{C},\theta,\vartheta)$, where
$\mathcal{A}$ is a set of \atomic{}s and $\mathcal{C} = \{\chi=(s,s')
| s,s'\in\mathcal{A}\}$ is a set of connections where $\chi$ indicates
a connection between two \atomic{}s $s$ and $s'$. We define an \atomic
as a section of road between two intersections, such that it preserves
the road's curvature but does not contain turns or sharp curves. A
connection becomes an intersection on the road that connects two
\atomic{}s. Note that these connections may extend the same road or
may turn into another road. The turn angle associated with a
connection $\chi$ is given by the function $\theta(\chi)$ and the
\atomic{}'s curvature is given by the function $\vartheta(s)$ as
defined in Equation~\ref{eq:curve_score}. In this graph construction,
we represent each \atomic $s$ by a vertex $v\in{}V$ and each
connection $\chi$ by an edge $e\in{}E$. \Cref{fig:graph} shows an
example road network and the corresponding graph construction. A
default speed limit is assigned to each \atomic based on the road type
in OpenStreetMap. For example, a `motorway' symbolizes interstates in
the USA that have speed limits $\approx 65mph$. The length, speed
limit, and geographic coordinates of the \atomic $s$ are stored as
attributes of the corresponding vertex $v$. The length and speed limit
are used to calculate the fastest time of travel between the end
points. \textit{It is important to note that this is a one time
  initialization step for every geographic area}.\\

\subsubsection{Spoofed Paths Generator}
\label{sec:algo1}

\begin{algorithm}[t]
\SetKwProg{Fn}{function}{:}{}
\SetKwFunction{spoofedpathsrec}{GenerateSpoofedPaths}

\DontPrintSemicolon
\KwIn{$G=(V,E)$, $Loc(s)$, $Loc(d)$, $N_P$}
\KwOut{$\mathcal{S} = \{p_1, \ldots, p_{N_P}\}$}

\BlankLine
$Initialization: \mathcal{S}\leftarrow{}\emptyset$; $p\leftarrow{}[\;]$; $v\leftarrow{}\emptyset$\\
$s\leftarrow{}getSourceVertex(Loc(s))$\\
$d\leftarrow{}getDestinationVertex(Loc(d))$\\
\spoofedpathsrec{$s$, $d$}\\
$\mathcal{S}\leftarrow{}selectTopPaths(\mathcal{S}, N_P)$\\

\BlankLine
\Fn{\spoofedpathsrec{$s$, $d$}} {
    $p\leftarrow{}p + [s]$\\
    $v\leftarrow{}v \cup \{s\}$\\
    
    \BlankLine
    \uIf{$s = d$}{
        $\mathcal{S}\leftarrow{}\mathcal{S} \cup \{p\}$\\
    }
    \uElse{
        \For{$e\in{}V$ such that $(s,e)\in{}E$}{
            \uIf{$e\not\in{}v$ {\bf and} $Filter(s, e, p)$ passed}{
                $p.score\leftarrow{}p.score * Score(s, e, p)$\\
                \spoofedpathsrec($e$, $d$)\\
            }
        }
    }

    \BlankLine
    $p\leftarrow{}p - [s]$\\
    $v\leftarrow{}v - \{s\}$\\    
}
\vspace{1mm}
\caption{Spoofed Paths Algorithm}
\label{alg:spoofed_paths}
\end{algorithm}

\begin{figure}[t]
	\centering
	\begin{subfigure}{.5\linewidth}
        \centering
		\includegraphics[width=\textwidth]{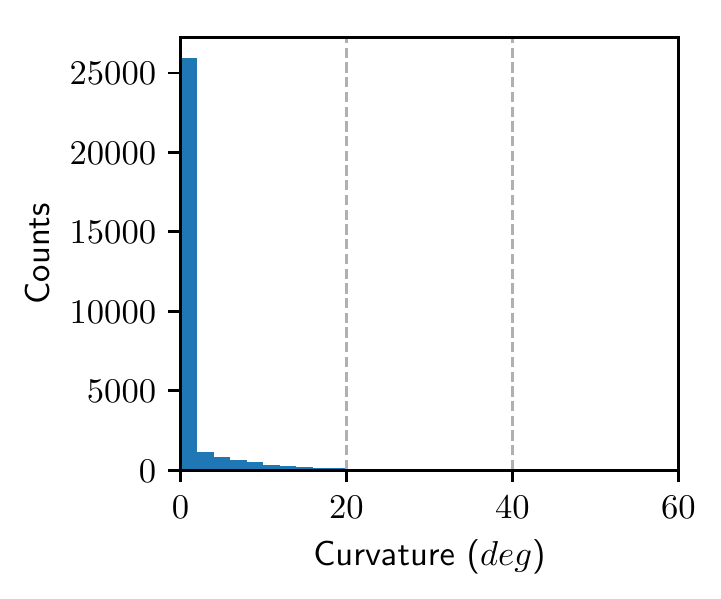}
		\caption{\label{fig:manhattan_curvature} Curvature Distribution}
    \end{subfigure}%
    \begin{subfigure}{.5\linewidth}
        \centering
		\includegraphics[width=\textwidth]{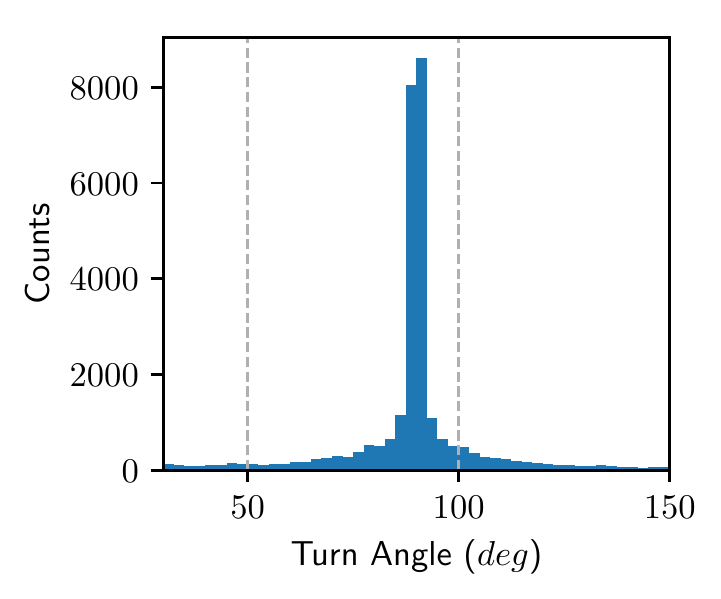}
		\caption{\label{fig:manhattan_turns} Turn Distribution}
    \end{subfigure}
	\caption{Curvature and Turn Distribution for Manhattan.}
	\label{fig:manhattan_distribution}
\end{figure}

Recall that the spoofed paths generator searches and compiles possible
paths between the source and destination points assigned to a specific
trip. We define spoofed paths as follows. \emph{The spoofed paths are
  a set of $N$ routes $\mathcal{S}$ = \{$\mathcal{S}_1, \ldots,
  \mathcal{S}_N$\} such that $\mathcal{S}_i$ has a higher likelihood
  of spoofing than $\mathcal{S}_j$, where $i < j$ and ${\mathcal{S}_i,
    \mathcal{S}_j} \in \mathcal{S}$. Each route $\mathcal{S}_i$
  contains a list of geographic coordinates starting and ending at the
  input source and destination.} Given the geographic area of the
attacker, the algorithm generates paths that maximize the probability
of finding similar road curvature and turn angles in other sections of
the area. Therefore, it maximizes the number of escape paths. It
leverages the fact that urban areas have regular patterns where most
roads typically run straight and turn angles are at right angles. This
is achieved by implementing a scoring scheme that ranks paths
containing such regular patterns higher than other non-regular paths
between the same source and
destination. \Cref{fig:manhattan_distribution} shows the curvature and
turn angle distribution for Manhattan and provides an intuition for
our approach. Here we see that most turn angles are $90\degree$ which
implies that given a path with all $\approx 90\degree$ turns, the
probability of finding another path with similar turn angles (i.e.,
all $\approx 90\degree$) will be high.

The idea underlying the spoofed paths generator is to find paths that
contain attributes likely to be found in other sections of the
graph. When such paths are found, they increase the likelihood of
finding similar paths to other destinations in the graph. To this
extent, we implement a scoring scheme that analyses the road curvature
and turn angles of the geographic area and maximizes the score of
paths that contain curvature and turns having a higher probability of
occurrence. The path search algorithm is implemented as a modified
Depth First Search (DFS) algorithm. A typical DFS implementation
computes a single path between a given source and destination. This
limits an attacker's ability to generate multiple spoofed paths
between these end points. We extend the basic DFS algorithm to compute
all \textit{plausible} non-cyclic paths between the source and
destination. For large graphs (typical for large cities), the above
modification results in an inefficient search where each vertex may be
visited numerous times. To scale the algorithm, we incorporate
filtering and scoring functions in order to speed up computation by
filtering out unlikely paths and pruning low scoring paths at every
iteration.

The spoofed paths generator algorithm (\Cref{alg:spoofed_paths}) takes
as input a graph $G=(V,E)$, the source $Loc(s)$ and destination
$Loc(d)$ geographic coordinates, and a count of output paths
$N_P$. The algorithm outputs a set of spoofed paths $\mathcal{S}$
sorted by the path score. The algorithm starts by initializing the
current path $p$ and a set of visited vertices $v$ (line 1). It uses
the attacker's source $s$ and destination $d$ vertices as parameters
to \spoofedpathsrec to recursively compute the output paths (lines 2
-- 4). In the end, these paths are sorted by score and the top $N_P$
paths are saved as the final set of spoofed paths $\mathcal{S}$ (line
5). Inside the \spoofedpathsrec function, the algorithm adds the
vertex $s$ to the current path $p$ and the visited set $v$ (lines 7 --
8) and adds this path $p$ to the output set $\mathcal{S}$ when the
destination vertex is found (lines 9 -- 10). Otherwise, the algorithm
traverses over the path's outgoing edges $e$ such that
$(s,e)\in{}E$. During this traversal (lines 12 -- 16), filtering is
applied to prune edges that are unlikely to occur (line 13) and a
scoring function is applied to rank remaining edges (line 14). The
filtering and scoring methodology are described next. The
\spoofedpathsrec function is recursively invoked for each outgoing
edge $e$ (line 15). Note that, in the end, the source $s$ vertex is
removed from the current path $p$ and visited set $v$ to backtrack and
proceed with the depth-first search (lines 17 -- 18).

\vspace{1mm}
\noindent \textbf{Scoring:} Recall that all the vertices of the graph
$G=(V,E)$ are \atomic{}s, and the edges connect two \atomic{}s
(c.f. \Cref{sec:graph}). The turn angle of an edge $\chi = (s, s')$,
where $(s, s') \in E$, is given by the function $\theta(\chi)$ and the
curvature of an \atomic $s$ is given by the function
$\vartheta(s)$. This curvature $\vartheta(s)$ can be computed from the
geographic coordinates of the \atomic. Let $\mathcal{B} =
\{\mathcal{B}_1, \ldots, \mathcal{B}_N\}$ denote the set of bearings
computed from $N$ geographic coordinates. Let $\mathcal{B}_0$ be the
bearing of an imaginary line connecting the first and last geographic
coordinates of this \atomic. The curvature $\vartheta(s)$ of this
\atomic is calculated as the normalized absolute difference of all
bearings in $\mathcal{B}$ from the reference bearing $\mathcal{B}_0$,
i.e.,
\begin{equation}\label{eq:curve_score}
\vartheta(s) = \frac{\sum_{i=1}^{N} |\mathcal{B}_i - \mathcal{B}_0|}{N}.
\end{equation}

The set of all road curvatures $\vartheta = \{\vartheta(s') |
\forall{}s' \in V\}$ and turn angles $\theta = \{\theta(\chi') |
\forall{}\chi' \in E\}$ represents the road structure of the
geographic area. \Cref{fig:manhattan_distribution} shows these
attributes for Manhattan. Note that most of the calculated curvature
values are $0\degree$ and most turn angles are at $90\degree$. This is
typical of Manhattan and other cities synonymous with grid-like road
structures. To use this information for scoring, a probability
distribution table is precomputed for the area. This table can be
represented as $P(\mathcal{G}) = \{P(c, t) | c \in \vartheta, t \in
\theta\}$, where each entry is the probability of occurrence of a
specific curvature and turn combination (rounded to the nearest
integer).

A path on the graph with $M$ vertices can be represented using each
vertex's curvature and the next edge's turn angle, i.e., $p = [(c_1,
  t_1), \ldots, (c_{M-1}, t_{M-1}), (c_M, 0)]$, where $c_i \in
\vartheta$ and $t_i \in \theta$. In the beginning, the path is
initialized to a score of $1$. For each vertex $\hat{s}$ and edge
$\hat{\chi} = (\hat{s}, s')$ added to the path, the probability
$P(\vartheta(\hat{s}), \theta(\hat{\chi}))$ is obtained from the table
$P(\mathcal{G})$. Note that, owing to the algorithm construction, all
connecting edges have equal probability of occurrence and are
independent of the current path. Therefore, the score at each vertex
is multiplied with the previous path score to calculate the compound
probability of all vertices in the path. The final path score is
calculated as
\begin{equation}\label{eq:path_score}
score = \prod_{i=1}^{M} P(\vartheta(s_i), \theta(\chi_i)).
\end{equation}

\vspace{1mm}
\noindent \textbf{Filtering:} The algorithm is designed to generate
all paths between the input source and destination. For a large graph,
the number of possibilities can be in the order of billions making
this search very inefficient. To scale the computation, the algorithm
uses the following filters to speed-up the search of plausible paths,
while enabling ranking. Given the current path $p$, source $s$, edge
$e$ and destination $d$, the algorithm filters the edge when the
path's distance summed with the euclidean distance between the edge
and destination exceeds a maximum allowed distance, i.e.,
$\textit{d}(p) + \textit{d}(c,d) > F * \textit{d}(\mathcal{P}_I)$
where $\textit{d}(.)$ denotes the distance of a path and
$\mathcal{P}_I$ denotes the shortest time path between the source and
destination. For this work, we set $F=1.2$ to only allow paths that
are similar in distance to the computed shortest path. The algorithm
also maintains the best $N$ paths at all times, and any new path $p'$
having a worse score is filtered. For our evaluation, we chose $N
=100$ in order to determine the attack efficiency in many cities for
many paths (the algorithm runs in around 1 minute for each
source/destination pair). However, a determined attacker with
sufficient resources can easily use a larger $N$ to increase the count
of spoofed paths. Furthermore, the adversary will only be interested
in a single source/destination pair of locations on each instance of
the attack, and can therefore take more time to derive the largest set
possible of spoofed and escape paths. The shortest path
$\mathcal{P}_I$ is also bounded by a rectangle (with added padding of
$m=1000$ meters) such that all edges outside the rectangle become out
of scope. Note that the above algorithm parameters are tunable and set
to conservative values in this work. We believe that the attack
performance can substantially improve when these parameters are tuned
more aggressively, e.g., setting $F=1.5$ and $N=1000$ (large values of
$N$ are very reasonable when focusing on a single
source/destination).\\

\subsubsection{Escape Paths Generator}
\label{sec:algo2}




\begin{algorithm}[t]
\SetKwProg{Fn}{function}{:}{}
\SetKwFunction{escapepathsrec}{GenerateEscapePaths}

\DontPrintSemicolon
\KwIn{$G=(V,E)$, $\mathcal{S}_I$}
\KwOut{$N_P$, $\mathcal{E} = \{p_1, \ldots, p_{N_P}\}$}

\BlankLine
$Initialization: \mathcal{E}\leftarrow{}\emptyset$; $N_P\leftarrow{}0$; $p\leftarrow{}[\;]$; $v\leftarrow{}\emptyset$\\
$s\leftarrow{}getSourceVertex(\mathcal{S}_I)$\\
$t\leftarrow{}getTurnsCount(\mathcal{S}_I)$\\
\escapepathsrec{$s$, $t$}\\

\BlankLine
\Fn{\escapepathsrec{$s$, $t$}} {
    $p\leftarrow{}p + [s]$\\
    $v\leftarrow{}v \cup \{s\}$\\
    
    \BlankLine
    \uIf{$len(p.turns) > t$}{
        \Return
    }

    \BlankLine
    \uIf{$len(p.turns) = t$}{
        $\mathcal{E}\leftarrow{}\mathcal{E} \cup \{p\}$\\
        $N_P\leftarrow{}N_P + 1$
    }
    
    \BlankLine
    \For{$e\in{}V$ such that $(s,e)\in{}E$}{
        \uIf{$e\not\in{}v$ {\bf and} $Filter(s, e, p, \mathcal{S}_I)$ passed}{
            $p.curve\leftarrow{}updateCurvature(s, e, p)$\\
            $p.turns\leftarrow{}updateTurns(s, e, p)$\\
            $p.score\leftarrow{}p.score * Score(s, e, p, \mathcal{S}_I)$\\
            \escapepathsrec($c$, $t$)\\
        }
    }

    \BlankLine
    $p\leftarrow{}p - [s]$\\
    $v\leftarrow{}v - \{s\}$\\    
}
\vspace{1mm}
\caption{Escape Paths Algorithm}
\label{alg:escape_paths}
\end{algorithm}

The idea behind the escape paths generator is to find all the paths an
attacker can travel to reach different destinations without raising
any alarms. An important consideration for this algorithm is that all
computed paths must have similar accelerometer and gyroscope patterns
to the spoofed paths, to avoid detection by GPS/INS tracking
systems. We formally define escape paths as follows. \emph{The escape
  paths corresponding to a spoofed path $\mathcal{S}_i$ are a set of
  $M$ routes $\mathcal{E}_i$ = \{$\mathcal{E}_{i_1}, \ldots,
  \mathcal{E}_{i_M}$\} such that $\mathcal{E}_{i_j} \ne
  \mathcal{S}_i$, but semantically similar to $\mathcal{S}_i$, for any
  $\mathcal{E}_{i_j} \in \mathcal{E}_i$. The paths are semantically
  similar when they have similar distances, road curvature and turn
  angles. These paths start at the input source, however, end at
  different destinations from the intended destination.}

Given a spoofed path, the escape paths algorithm
(\Cref{alg:escape_paths}) generates a set of escape paths with similar
distances, road curvatures and turn angles to the spoofed path. The
algorithm is similar to that of the spoofed paths generator. The main
differences being that the algorithm uses each spoofed path
$\mathcal{S}_I$ generated in the previous stage as input, where
$\mathcal{S}_I \in \mathcal{S}$, and outputs a set of escape paths
$\mathcal{E}$. Furthermore, the escape paths generator algorithm uses
the count of turns in the spoofed path as a parameter to
\escapepathsrec (lines 3 -- 4) and checks whether the desired count of
turns has been reached for the escape path under consideration (lines
10 -- 12).

The deviations from the spoofed paths (to avoid INS detection) can be
determined by analyzing the noise sensitivity of the inertial sensors
used for tracking.  We demonstrate that commodity accelerometers and
gyroscopes present challenges in accurately calculating the distances,
road curvature and turn angles which can allow an attacker to travel
to multiple destinations without detection. We also show that
magnetometers can be easily spoofed rendering them incapable of
detecting anomalies in the heading direction of the vehicle. Our
analysis of the accelerometer and gyroscope noise and the potential of
magnetometer spoofing are reported in \Cref{sec:sensors_eval}. Unlike
the spoofed paths generator algorithm that ranked paths by score, the
escape paths computed by this algorithm always have a score of
$1$. The intuition is that all paths that pass the algorithm's filters
are certain to avoid detection by INS tracking systems.





\vspace{1mm}
\noindent \textbf{Filtering:} In this algorithm, we represent the
input spoofed path by $\mathcal{S}_I = \{(d_I, \vartheta_I,
\theta_I)\}$ where $d_I$ and $\vartheta_I$ denote the set of distances
and road curvatures between intersections and $\theta_I$ denotes the
turn angles at the intersections. We first present the idea of
filtering using just turn angles $\theta_I$, and later expand the
discussion to include distances $d_I$ and road curvatures
$\vartheta_I$. Let $\theta_I = \{\theta(\chi_1), \ldots,
\theta(\chi_K)\}$ be the derived turn angles of the spoofed path,
where $K$ is the number of intersections. A turning connection $\chi'
= (s, e)$ in the escape path, where $(s, e) \in E$, is valid for an
intersection $k \in K$ when the turn angle difference is below a set
threshold value $\mathcal{T}_\theta$, i.e., $|\theta(\chi_k) -
\theta(\chi')| \le \mathcal{T}_\theta$. The parameter
$\mathcal{T}_\theta$ depends on the noise sensitivity of the gyroscope
sensor.

The filter for distances $d_I$ is similar to turn angles. Let $d_I =
\{d_1, \ldots, d_{K+1}\}$ be the derived distances of the spoofed path
traveled between $K$ intersections. For an intersection $k \in K$,
$d_k$ represents the path's distance from the previous intersection
$k-1$, i.e., $d_k = \textit{d}(k) - \textit{d}(k-1)$ where
$\textit{d}(.)$ denotes the total distance of the spoofed path at a
given intersection. Note that $k=0$ is the source of the path and
$k=K+1$ is the destination of the path. A connection $\chi'$ in the
escape path is valid for intersection $k$ when its path distance from
previous intersection $k-1$ is between a range defined by the $k^{th}$
intersection of the spoofed path, i.e., $d_k * T_{d1} \le
\textit{d'}(k) - \textit{d'}(k-1) \le d_k * T_{d2}$. Here,
$\textit{d'}(.)$ denotes the distance of the escape path at an
intersection. The above parameters $T_{d1}$ and $T_{d2}$ depend on the
noise sensitivity of the accelerometer sensor.

The filter for road curvature $\vartheta_I$ is more complex than turn
angles and distances. The reason is that, given an intersection $k \in
K$, the distance $d_k$ and turn angle $\theta(\chi_k)$ are scalars
while $\vartheta(s_k)$ is a vector that must be derived from bearings
of the road segment $s_k$ between intersections $k-1$ and $k$. Two
different vectors of bearings $\mathcal{B}_k$ and $\mathcal{B}'$ for
road segments $s_k$ and $s'$, respectively, cannot be compared
directly as they may be of different lengths and in different
orientations, e.g., $\mathcal{B}_k$ may be directed north when
$\mathcal{B}'$ is directed east. Our idea of calculating the road
curvature similarity, denoted by $\mathcal{C}(s_k, s')$, is to
translate these bearings to the same size $N$ using linear
interpolation, convert the interpolated bearings to curvature, and
then compare the curvatures. Let $\mathcal{B}_{Ik}$ and
$\mathcal{B}'_I$ represent the interpolated bearings for
$\mathcal{B}_k$ and $\mathcal{B}'$, respectively. The curvature of a
road segment $s$ with $M$ bearings $\mathcal{B} = [b_1, \ldots, b_M]$
can be derived by subtracting the first bearing $b_1$ from all the
bearings in $\mathcal{B}$, i.e., $\vartheta(s) = [(b_1-b_1), \ldots,
  (b_M-b_1)]$. Let $\vartheta(s_k)$ and $\vartheta(s')$ be the
curvatures derived from $\mathcal{B}_{Ik}$ and $\mathcal{B}'_I$,
respectively. The curvature similarity of the two segments can then be
represented as:
\begin{equation}\label{eq:curve_similarity}
\mathcal{C}(s_k, s') = \{|c_k-c'| \quad \forall{}c_k \in \vartheta(s_k), \forall{}c' \in \vartheta(s')\}.
\end{equation}
A connection $\chi'$ in the escape path is valid for intersection $k$ when the maximum curvature similarity value is below a set threshold value $\mathcal{T}_\vartheta$, i.e., $max(\mathcal{C}(s_k, s')) \le \mathcal{T}_\vartheta$. Like turn filtering, this parameter $\mathcal{T}_\vartheta$ also depends on the gyroscope noise sensitivity.

To avoid detection, the above discussed constraints must hold for all
$K$ intersections of the escape path. Therefore, a escape path is
considered valid if and only if all the following conditions are met.

\[\small
\begin{array}{l@{\hspace{2em}}l}
|\theta(\chi_k) - \theta(\chi')| \le \mathcal{T}_\theta, \quad &  \forall \: k = 1, \ldots, K  \\
d_k * T_{d1} \le \textit{d'}(k) - \textit{d'}(k-1) \le d_k * T_{d2}, & \forall\: k = 1, \ldots, K+1  \\
max(\mathcal{C}(s_k, s')) \le \mathcal{T}_\vartheta, & \forall\: k = 1, \ldots, K+1 
\end{array}
\]
\vspace{2mm}

\section{Attack Impact: Implementation and Evaluation}

In this section, we present the implementation of our attack and evaluate evaluate its effectiveness in various cities across the globe. First, we evaluate the accuracy of inertial sensors and derive realistic noise threshold settings for \suitename algorithm. Then, we describe the details of our experimental setup and the methodology. Finally, we present the results of our evaluation using two metrics, (i) displacement from the assigned destination and (ii) coverage area of the escape paths. 



\subsection{Accuracy of Inertial Sensors}
\label{sec:sensors_eval}

The sensor data for evaluating the noise sensitivity of accelerometers
and gyroscopes was obtained from an open
dataset~\cite{narain_snp16}. This dataset comprises of accelerometer,
gyroscope and magnetometer samples recorded from $\approx 140$ real
driving experiments in the cities of Boston and Waltham, MA, USA. The
sensor samples were collected on $4$ smart phones (HTC One M7, LG
Nexus 5, LG Nexus 5X, and Samsung S6). The GPS traces for these routes
were also recorded for ground truth comparison. The authors of that
work focused specifically on gyroscope noise during turns. We extend
their work to also determine noise sensitivity when distance is
calculated from the accelerometer sensor, as well as when road
curvature is calculated from the gyroscope sensor.

\begin{figure}[t]
	\centering
	\begin{subfigure}{.5\linewidth}
        \centering
		\includegraphics[width=\textwidth]{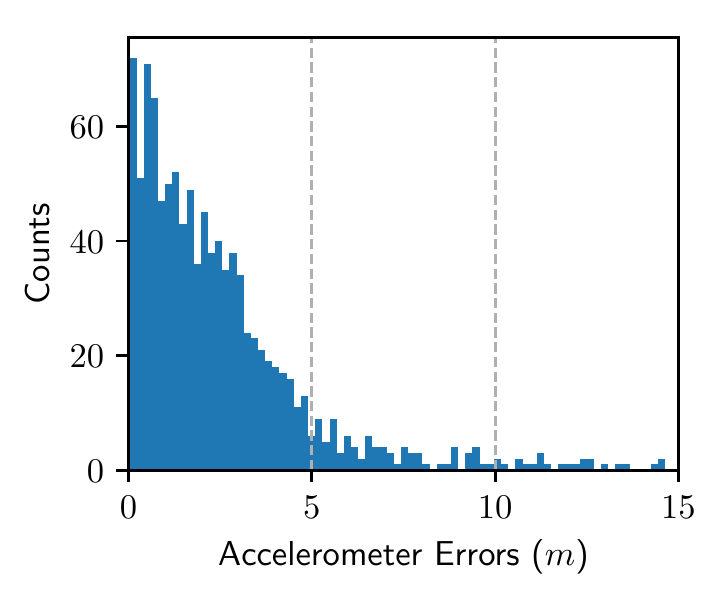}
		\caption{\label{fig:distance_accuracy} Distance calculation errors}
    \end{subfigure}%
    \begin{subfigure}{.5\linewidth}
        \centering
		\includegraphics[width=\textwidth]{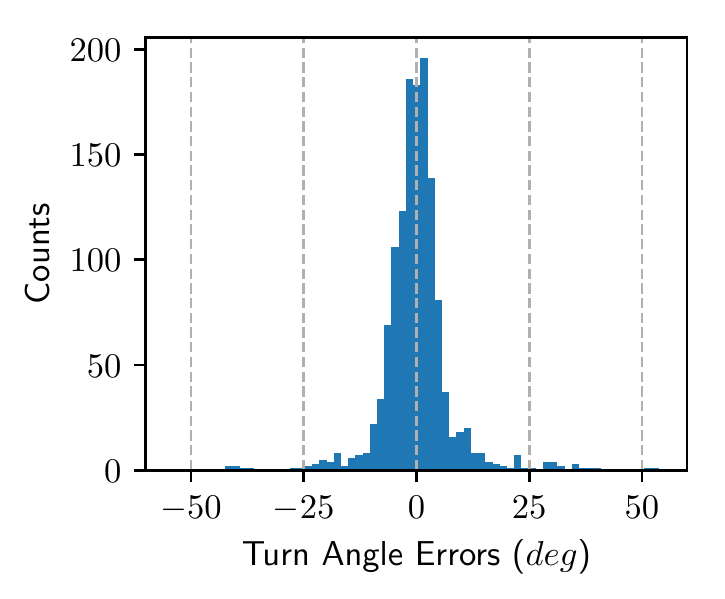}
		\caption{\label{fig:turns_accuracy} Turn angle errors}
    \end{subfigure}
    \begin{subfigure}{.5\linewidth}
        \centering
		\includegraphics[width=\textwidth]{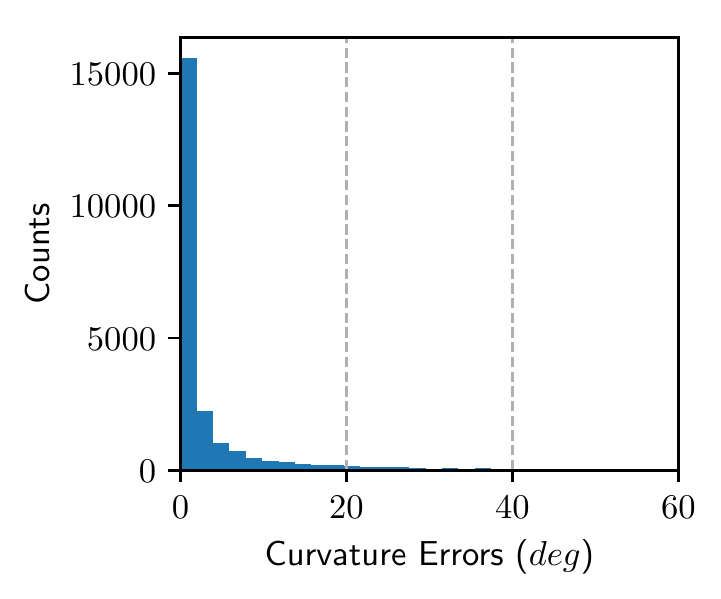}
		\caption{\label{fig:curve_accuracy} Curvature errors}
    \end{subfigure}
	\caption{Accelerometer (distance) and Gyroscope (road curvature and turn angle) errors measured using real experiments.}
	\label{fig:sensors_accuracy}
\end{figure}

\subsubsection{Accelerometer Accuracy}

The accelerometer sensor can be used to calculate the distance
traveled for a path. This data can be represented as a vector $a =
[(a_1+n_1), \ldots, (a_T+n_T)]$ sampled at discrete time intervals $t
\in T$, where $a_t$ is the true acceleration experienced by the device
on the $x$, $y$ and $z$ axis, and $n_t$ is an unknown noise quantity
caused by several factors. For example, the sensors have an inherent
bias due to manufacturing defects such as axis misalignment. Another
source of noise is the vibrations caused by the mechanical structure
of the vehicle and the engine. Additional noise is induced on the
sensor due to external environments such as road conditions and
traffic.

We are interested in finding the range of divergence from the actual
values due to $n_t$, when distance is calculated from the
accelerometer data. To obtain this range, we calculated the distances
between intersections using accelerometer data for each sensor path in
the data-set, and compared it to the actual distances obtained from
OpenStreetMap. Note that, to reduce the impact of noise, we performed
the calibration and rotation techniques described
in~\cite{narain_snp16} before calculation. We also average multiple
samples together to further reduce the impact from noise. As distances
may significantly vary between intersections, we represent the
distance error as a ratio of the derived accelerometer distances to
the actual distances. More precisely, if $d_s$ is a vector of $N$
derived accelerometer distances and $d_a$ is a vector of $N$ actual
distances, then the errors $e_a$ can be represented as a vector $e_a =
[(d_{s_1} / d_{a_1}), \ldots, (d_{s_N} /
  d_{a_N})]$. \Cref{fig:distance_accuracy} shows the distribution of
the errors $e_a$. Note that the desired value for an error should be
near $1$, however, we see large variations ranging between $0.1$ to
$5$. This indicates that the accelerometer sensor is unsuitable for
distance calculation and enables an attacker to travel much larger
distances than the intended path. Recall that the escape paths
generator algorithm uses parameters $T_{d1}$ and $T_{d2}$ to filter
connections of the escape paths based on distances
(\Cref{sec:algo2}). These parameters are chosen from the error
distribution $e_a$ such that the allowed range is based on the
$75^{th}$ percentile of the distribution, i.e., $T_{d1}=0.2$ and
$T_{d2}=3.3$.

\subsubsection{Gyroscope Accuracy}

The gyroscope sensor can be used to measure the turn angles and the
road curvature of the path. This data can also be represented as the
vector $g = [(g_1+n_1), \ldots, (g_T+n_T)]$, where $g_t$ is the rate
of angular change experienced by the device on the $x$, $y$ and $z$
axis, and $n_t$ is an unknown noise quantity. In this case, however,
the impact of $n_t$ is not as significant as accelerometers and the
measurements are closer to the actual values.

We are interested in finding the turn angle errors and the curvature
errors calculated from the gyroscope data, in comparison to the actual
values derived from OpenStreetMap. To calculate the turn errors, we
use a similar approach to~\cite{narain_snp16} in that we define a turn
error as the absolute difference between the gyroscope derived turn
angle and the actual turn angle. However, we are interested in the
overall error distribution for all the phones instead of individual
phones. \Cref{fig:turns_accuracy} shows the distribution of the turn
angle errors for all the turns in the data-set. The distribution
reaffirms that the gyroscope is much more accurate than the
accelerometer where $75$\% of the turn errors are within $5.5\degree$.

To calculate the curvature errors, recall our technique for
calculating curve similarity $\mathcal{C}(s_k, s')$ for two road
segments $s_k$ and $s'$ between the $(k-1)^{th}$ and $k^{th}$
intersections (\Cref{eq:curve_similarity}). The road curvature
$\vartheta(s_k)$ is already known in the form of the gyroscope
data. However, this curvature must be interpolated to the same length
as $\vartheta(s')$. Given the union of curve similarity sets for all
$K$ intersections for $N$ sensor paths $\mathcal{C} =
\bigcup_{i=1}^{N} \mathcal{C}_i$, where $\mathcal{C}_i =
\bigcup_{j=1}^{K} \mathcal{C}(s_j, s'_j)$, the curvature errors $e_c$
is simply a set of absolute differences between all the points in the
two curves, i.e., $e_c = \{|c_s - c_a| \quad \forall{}[c_s, c_a] \in
\mathcal{C}\}$. \Cref{fig:curve_accuracy} shows the distribution of
the curve errors. Recall that the escape paths generator algorithm
defines parameters $T_\theta$ and $T_\vartheta$ to filter connections
based on turn angles and curvature, respectively
(\Cref{sec:algo2}). Based on the $75^{th}$ percentile of the error
distributions, we set the parameters to $T_\theta = 5.5\degree$ and
$T_\vartheta = 2.8\degree$ in our evaluations.

\subsubsection{Magnetometer Spoofing}





\begin{figure}[t]
	\centering
	\begin{subfigure}{\linewidth}
        \centering
		\includegraphics[width=\linewidth]{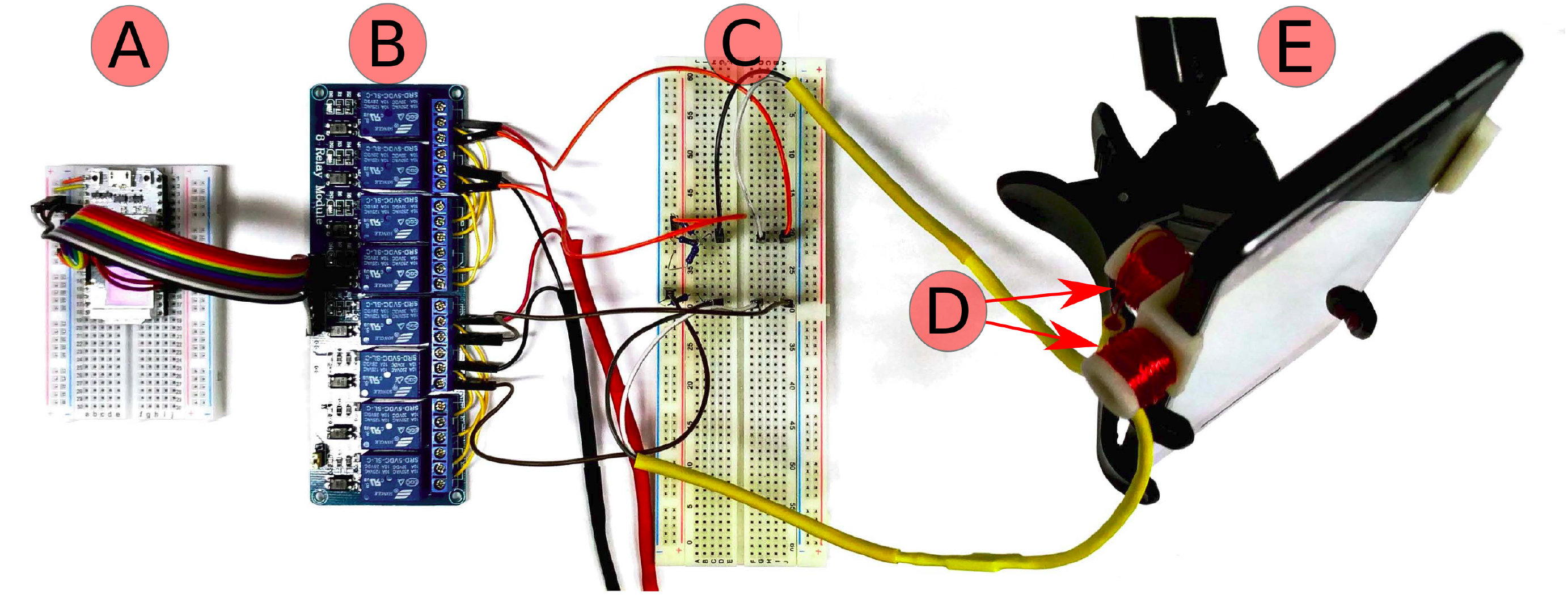}
		\caption{\label{fig:spoofing_setup} Experimental setup used for magnetometer spoofing}
    \end{subfigure}
    \begin{subfigure}{\linewidth}
        \centering
		\includegraphics[width=.15\linewidth]{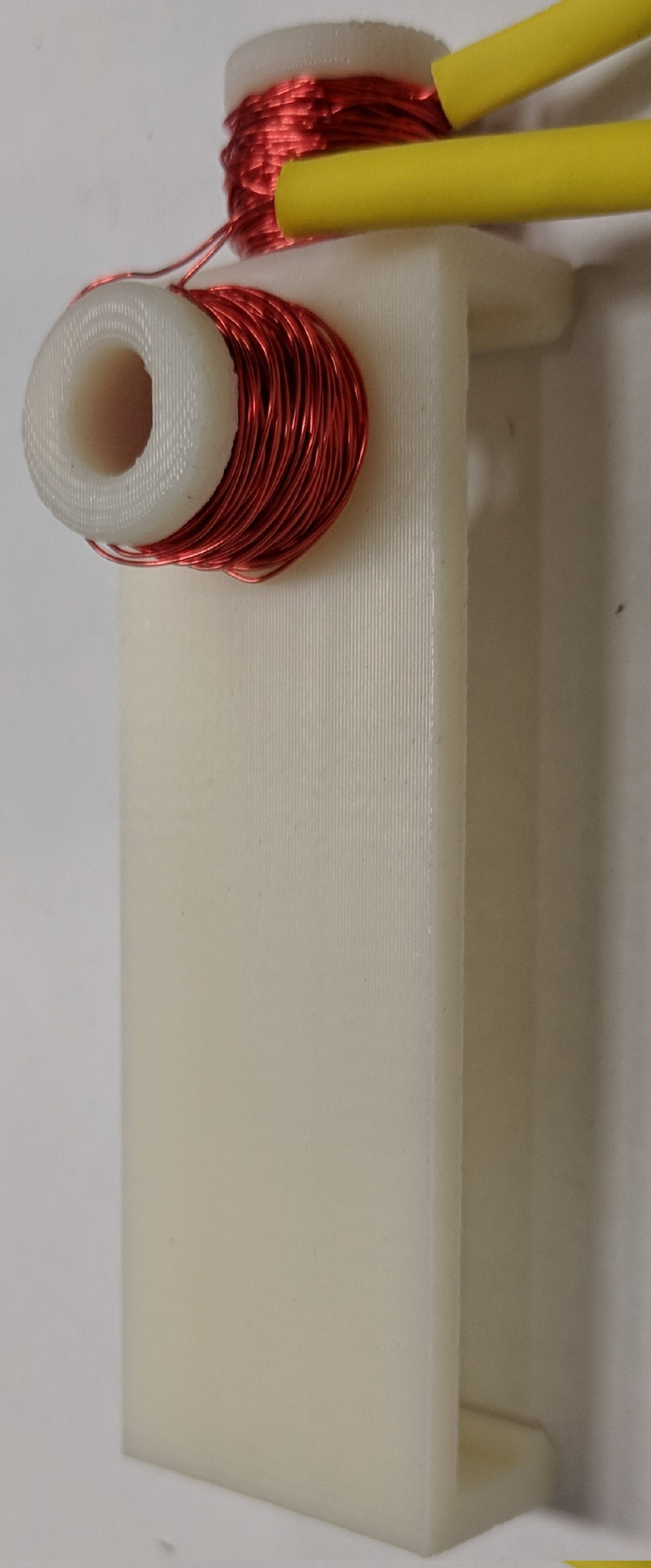}\hspace{2mm}
		\includegraphics[width=.565\linewidth]{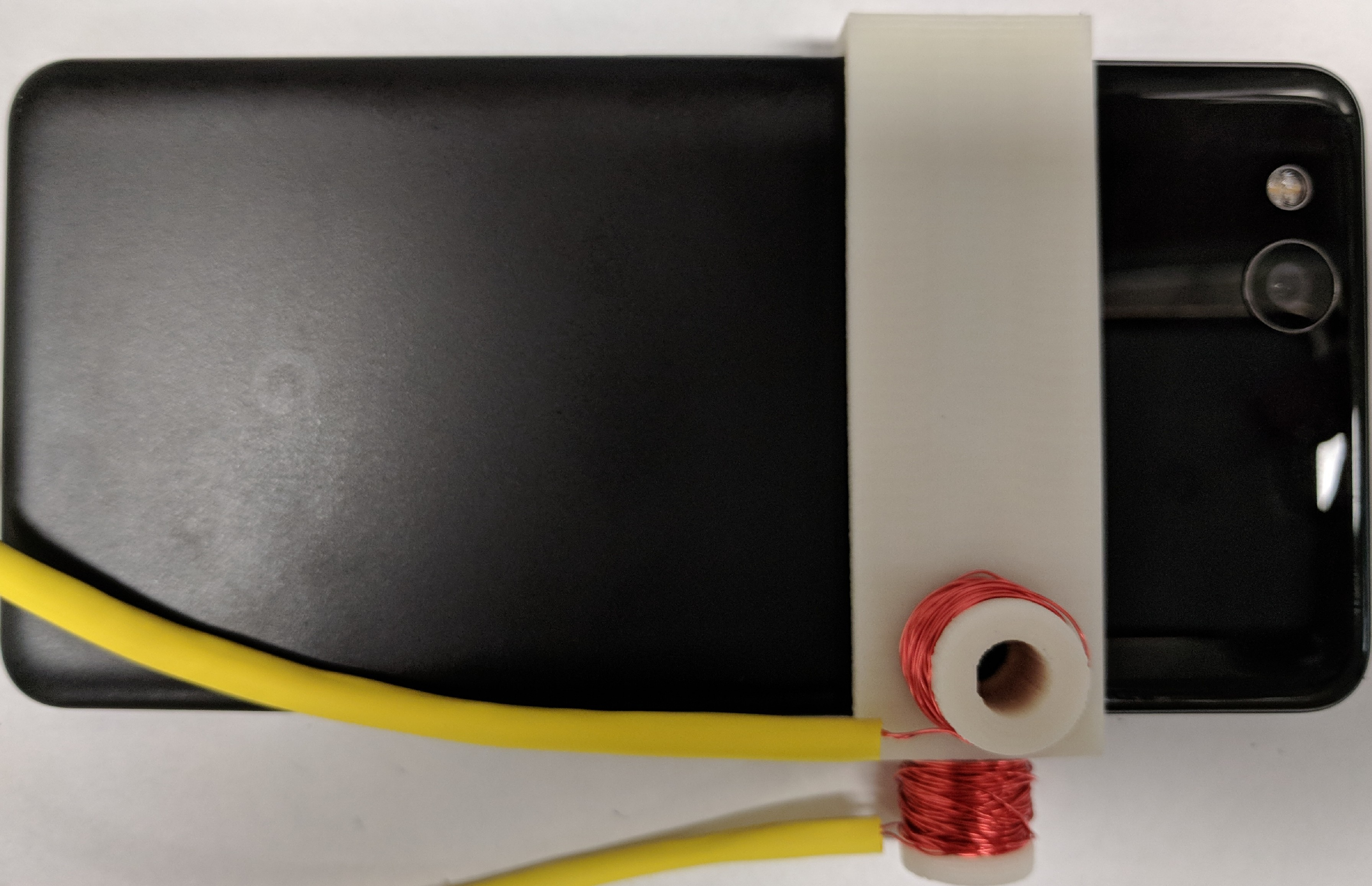}
		\caption{\label{fig:coil_phone} The two-coil system attached to a Google Pixel 2}
    \end{subfigure}
	\caption{The experimental setup implemented for demonstrating the potential of magnetometer spoofing.}
\end{figure}

\begin{figure}[t]
	\centering
	\begin{subfigure}{.51\linewidth}
        \centering
		\includegraphics[width=\textwidth]{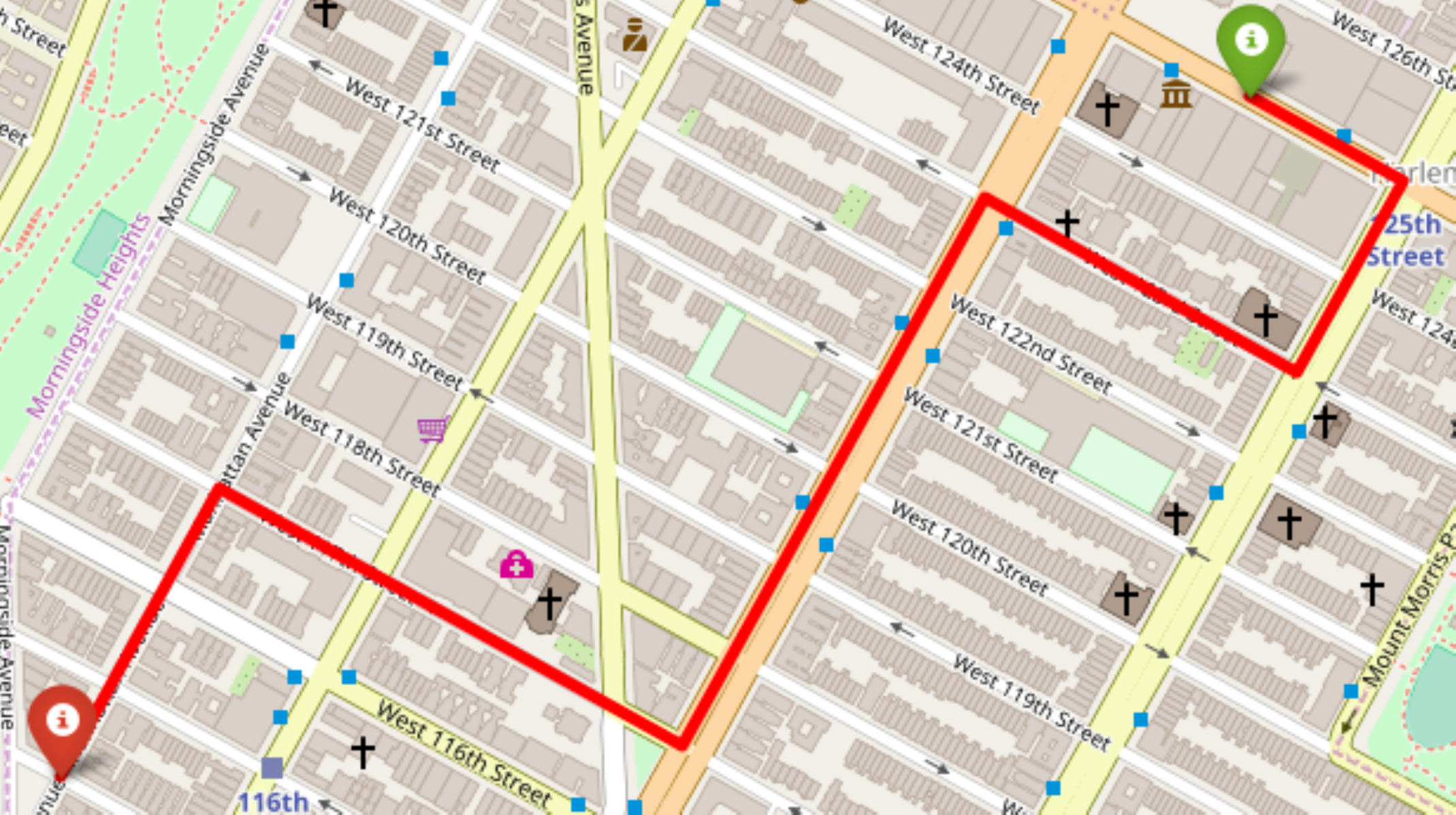}
		\caption{\label{fig:manhattan_path} Example Route in Manhattan}
    \end{subfigure}%
    \begin{subfigure}{.49\linewidth}
        \centering
		\includegraphics[width=\textwidth]{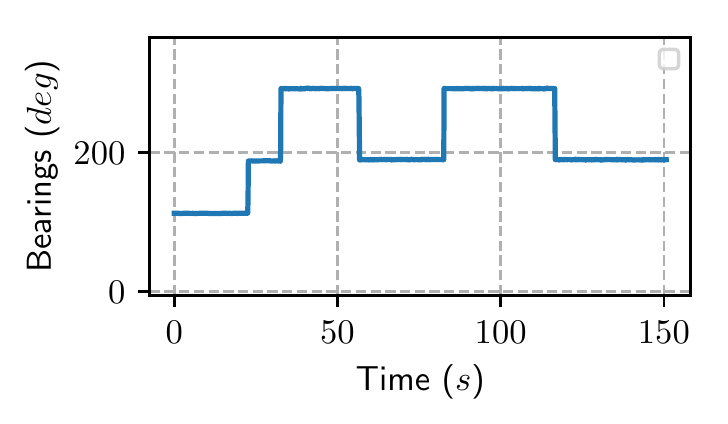}
		\caption{\label{fig:spoofed_bearings} Spoofed Bearings}
    \end{subfigure}
	\caption{An example of spoofing the magnetometer bearings for an example route in Manhattan.}
	\label{fig:manhattan_spoofing}
\end{figure}

As a proof of concept, we built a prototype of a magnetometer spoofer
for the Google Pixel 2 smart phone. Our experimental setup is shown
is \Cref{fig:spoofing_setup} and consists of the following modules:
(A) an ESP32 microcontroller, (B) a 8-channel relay module, (C)
resistors for controlling current flow, (D) a two coils system, and
(E) a Google Pixel 2 mounted on a car mount. We first identified the
exact location of the magnetometer which is on the top-left of the
phone ($42mm$ from the top and $7mm$ from left edge of the phone). We
designed and 3D printed a two-coils system, shown
in \Cref{fig:coil_phone}, that snaps on to the phone and allows the
wrapping of enameled magnet wire. We focused on controlling the $x$
and $y$ axes as they are easily reachable. Using two coils each
targeting one of the axes allows full control of the magnetic field in
a plane.  We used the following solenoid magnetic field formula to
estimate the intensity:
$$
B = k\mu_0 n I
$$
where $k$ is the relative permeability, $\mu_0 = 4\pi 10^{-7}$ H/m,
$n$ is the coil turn density, and $I$ is the electric current. Our
coils turn density $n$ is $155$ turns/meter since we used 5 layers of
$28$ AWG enameled magnet wire. Without a core ($k=1$), we estimated a
magnetic field of $98uT$ with a current of $5mA$, which is strong
enough to impact the magnetometer. Note that if the magnetometer is
not accessible in other systems, it is possible to use larger coils or
channel the magnetic field using materials with higher relative
permeability. While the relative permeability of air is $1$, it is
$5,000$ for iron, and $200,000$ for iron annealed in hydrogen. To
control the current in each of the coils, we used the ESP32
microcontroller (Heltec WiFi Kit 32) with a sufficient number of
GPIO/DAC pins to control the 8-channel relay module augmented with
variable resistors for current tuning. The spoofer was written in
Python and takes as input a sequence of bearings and durations. It
sets the current in the coils to trigger turns with a timing that
matches the input durations. The spoofing of an example route in
Manhattan is shown in \Cref{fig:manhattan_spoofing}.

\subsection{Simulation Setup and Evaluation Methodology}

\begin{figure}[t]
	\centering
	\begin{subfigure}{.5\linewidth}
        \centering
		\includegraphics[width=\textwidth]{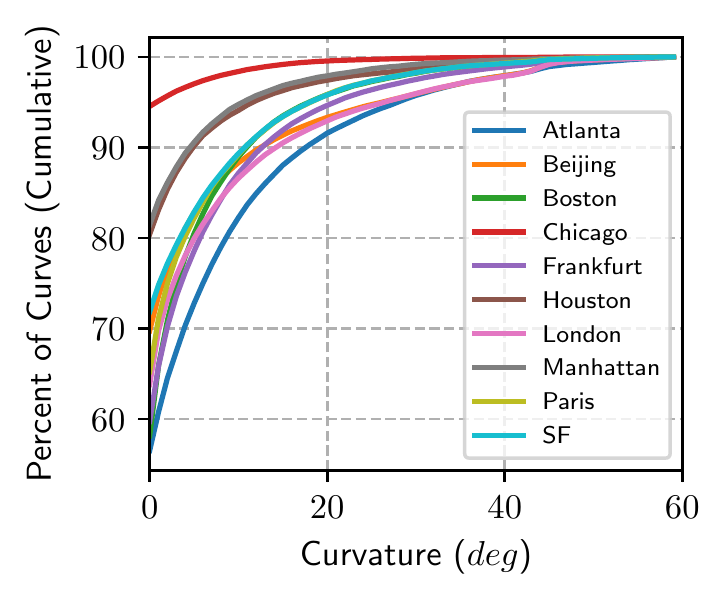}
		\caption{\label{fig:all_curvature} Curvature Distributions}
    \end{subfigure}%
    \begin{subfigure}{.5\linewidth}
        \centering
		\includegraphics[width=\textwidth]{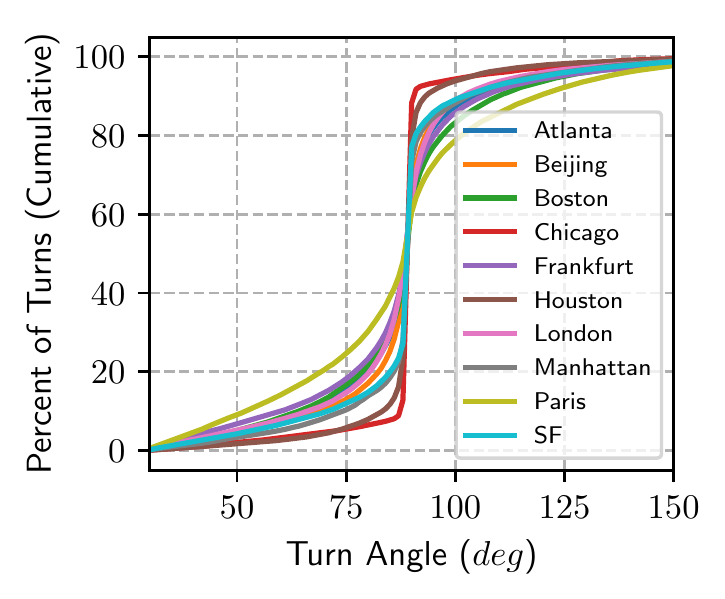}
		\caption{\label{fig:all_turns} Turn Distributions}
    \end{subfigure}
	\caption{Comparison of the Curvature and Turn Distribution for selected cities.}
	\label{fig:all_distributions}
\end{figure}

We implemented the \suitename attack algorithms in PyPy, a JIT compiler based alternative implementation of Python. We used two servers running Intel Xeon CPUs at $2.40{}GHz$ with $12$ cores and $20{}GB$ of RAM to execute the algorithms and evaluate its performance \ie how far can an attacker escape, given a start and end point, without being detected.\\ 

\noindent\subsubsection*{\textbf{Selection of cities}}
We evaluate the effectiveness of our attack on the road networks of 10 major cities across the globe. The following cities were chosen across the continents of North America, Europe and Asia for the evaluation: Atlanta, Boston, Chicago, Houston, Manhattan and San Francisco (North America), Beijing (Asia), London, Frankfurt and Paris (Europe). The cities were chosen to represent the entire spectrum of urban characteristics such as major logistics and transportations hubs, dense population, city planning (e.g., grid-like or circular), \etc \Cref{fig:all_distributions} shows the cumulative road curvature and turn distributions for all selected cities. Recall that the road curvatures are calculated using \Cref{eq:curve_score}. We can observe that Chicago and Manhattan have mostly straight roads and right angled turns while the road networks of London and Paris have very unique characteristics.\\

\noindent\subsubsection*{\textbf{Generation of spoofed and escape routes}}
The evaluation was performed by running simulations for every selected city. This simulation data comprised of $1000$ randomly generated paths in every city, such that the path distances were uniformly distributed between $1km$ and $21kms$. The intention was to evaluate the potential of spoofing also as a function of the path distance. The simulation paths were generated as follows: (i) a random `Home' and `Work' location were chosen from OpenStreetMap inside the interest area, (ii) the geographic coordinates of the end points were retrieved, and (iii) the coordinates were given as input to the attack algorithms to compute the spoofed and escape paths. Recall that the spoofed paths are all possible paths between the source and destination points assigned to a specific trip and escape paths are all the paths an attacker can travel to reach different destinations without being detected by the GPS/INS based monitoring system. A `Home' location can be chosen as a way or node in OpenStreetMap whose building type is one of the following: `apartments', `house', `residential', or `bungalow'. Similarly, a `Work' location can be chosen from the `commercial' or `industrial' tags.
 

%
\subsection{Evaluation Results}
We measure the performance of our attack using the two metrics: (i)
displacement from the actual destination and (ii) coverage area.

\subsubsection*{\textbf{Displacement from Intended Destination}}

\begin{figure}[t]
	\centering
	\begin{subfigure}{.24\textwidth}
        \centering
		\includegraphics[width=\textwidth]{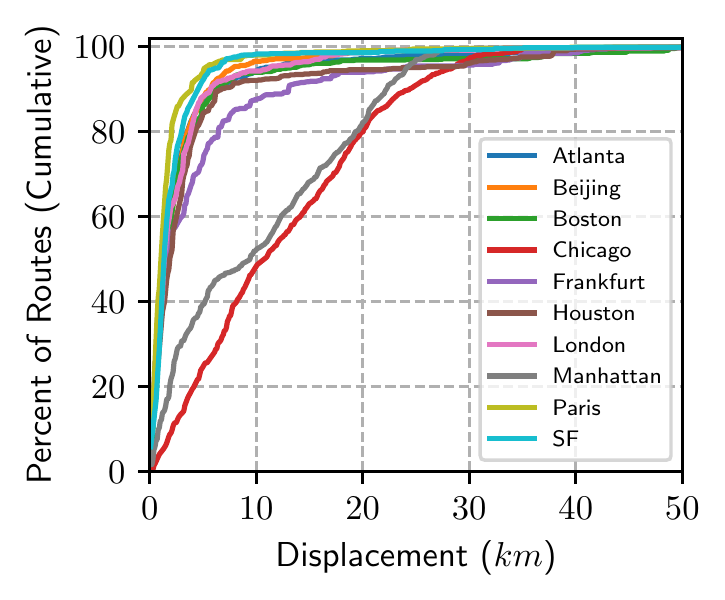}
		\caption{\label{subfig:max_displacement} An attacker's deviation from intended or assigned destination for the generated paths in all cities}
		
    \end{subfigure}\hfill
    \begin{subfigure}{.24\textwidth}
        \centering
		\includegraphics[width=\textwidth]{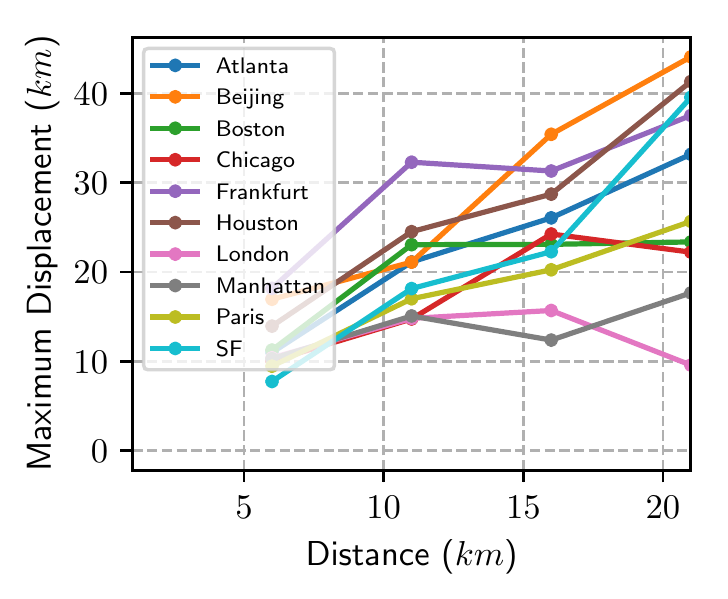}
		\caption{\label{subfig:displacements_cdf} The maximum displacement in every city for specific assigned path lengths}
    \end{subfigure}
    \caption{The displacement from intended destination and the maximum displacement in every city chosen for evaluation.}
	\label{fig:displacements}
\end{figure}

We define displacement from the intended destination as the farthest distance an attacker can reach for a chosen trip (i.e., given a start and end point) without being detected. For every evaluation route, escape and spoofed paths are generated as described previously. We then calculate the euclidean distance between the destinations an attacker reaches by taking the escape route and the actual intended destination \ie the assigned end point for the trip. We present our results in~\Cref{fig:displacements}.~\Cref{subfig:displacements_cdf} shows the attacker's deviation from the intended or assigned destination for the generated routes in all 10 cities. It can be observed that in majority of the cities, more than $20\%$ of the routes allow more than 10 km deviation from the intended destination. There are at least $10\%$ of the routes in all selected cities where the attacker is able to reach points as far as $30\unit{km}$ away from the assigned destination. Chicago and Manhattan perform the worst among the selected cities with more than $40\%$ of the routes allowing a displacement of $15\unit{km}$ or above. This is due to the regular patterns that exist in these cities' road network. Figure~\ref{subfig:max_displacement} shows the maximum displacement in each city for specific assigned route lengths. It is important to observe that in Manhattan and Chicago the maximum displacement caused is independent of the assigned route distance. This is due to the structure of the cities itself. For example, Manhattan is a narrow strip with grid like structures and therefore maximum displacement saturates at some point. However, for a city like Beijing there are routes that allow an attacker to spoof his location to as far as $40\unit{km}$ away from the intended location.\\




\subsubsection*{\textbf{Coverage Area of Spoofed Paths}}

The goal of this evaluation is to determine the percentage of area an
attacker can cover by traveling the escape paths generated for a given
source $Loc(s)$ and destination $Loc(d)$ geographic coordinates. Let
$A$ denote the total geographic area of interest to an attacker. For
this evaluation, we define this area as a circle of radius $r
= \textit{d}(Loc(s), Loc(d))$ with center at $Loc(s)$ where $r$ is the
euclidean distance between the source and destination. The above area
may comprise of water bodies which must be accounted for more accurate
coverage. Let $A_L$ denote the area of land within the interest
area. Within $A_L$, let $A_C$ denote the area that the attacker can
cover if he is willing to walk a small distance $r'$ from an escape
destination. The value $(A_C / A_L) * 100$ then expresses the
percentage of coverage area of the escape paths.

The area $A_L$ is not trivial to calculate as the location of water
bodies are not pre known within the interest area. The area $A_C$ is
also not trivial to calculate as the escape destinations may be
densely populated and many may overlap. To solve this, we implemented
Monte-Carlo simulations to estimate the above areas. The simulation
works by generating millions of uniformly distributed points within
the interest area. It maintains two separate counters: $P_L$ to count
all the points that are on land (i.e., within $r'$ meters of any
road), and $P_C$ to count all points within an escape destination's
radius (i.e., within $r'$ meters of any escape destination). With
these counters, the area $A_L$ can be calculated as $A_L = (P_L / P) *
A$, where $P$ is the total number of points, and the area $A_C$ can be
calculated as $A_C = (P_C / P) * A$. Therefore, the final percentage
of coverage area of the escape paths using Monte-Carlo simulation can
be expressed as $(P_C / P_L) * 100$. The percentage of coverage is the
ratio of the coverage area calculated (using a walking radius of 100
m) to the total area of land calculated using the Monte-Carlo
simulation.

\begin{figure*}[t]
	\centering
	\begin{subfigure}{.25\textwidth}
        \centering
		\includegraphics[width=\textwidth]{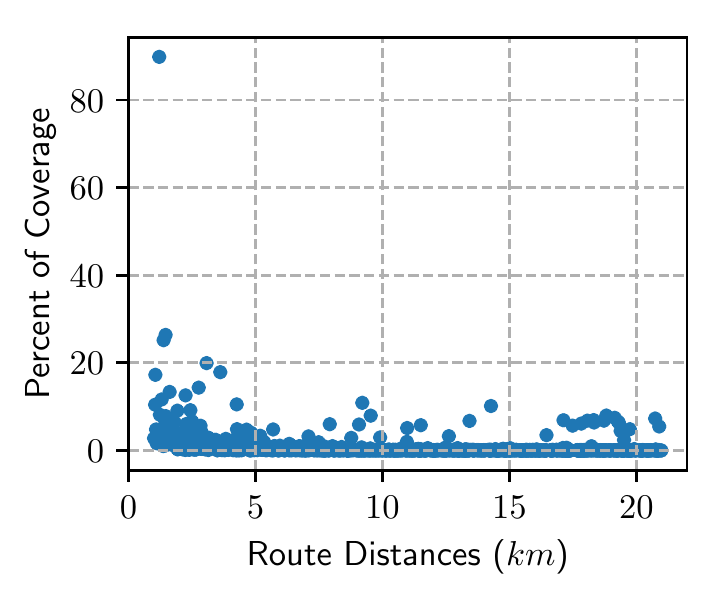}
		\caption{\label{fig:coverage_atlanta} Atlanta}
    \end{subfigure}%
    \begin{subfigure}{.25\textwidth}
        \centering
		\includegraphics[width=\textwidth]{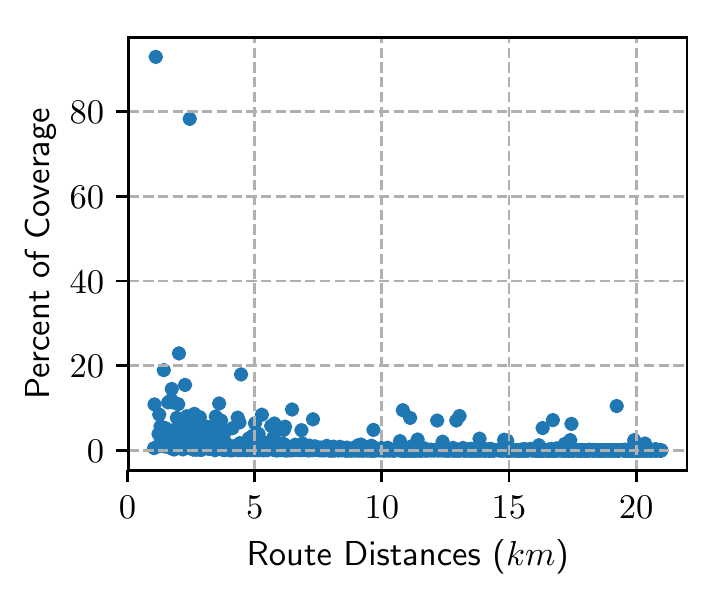}
		\caption{\label{fig:coverage_beijing} Beijing}
    \end{subfigure}%
    \begin{subfigure}{.25\textwidth}
        \centering
		\includegraphics[width=\textwidth]{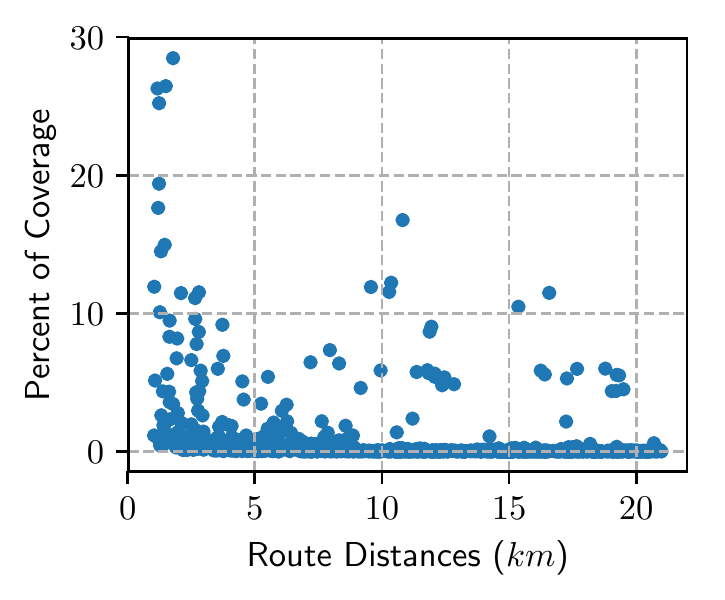}
		\caption{\label{fig:coverage_boston} Boston}
    \end{subfigure}%
    \begin{subfigure}{.25\textwidth}
        \centering
		\includegraphics[width=\textwidth]{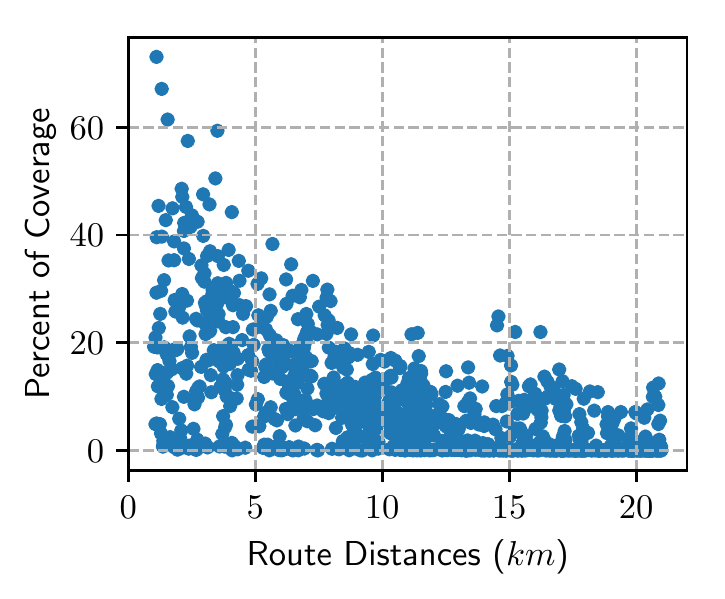}
		\caption{\label{fig:coverage_chicago} Chicago}
    \end{subfigure}
    \begin{subfigure}{.25\textwidth}
        \centering
		\includegraphics[width=\textwidth]{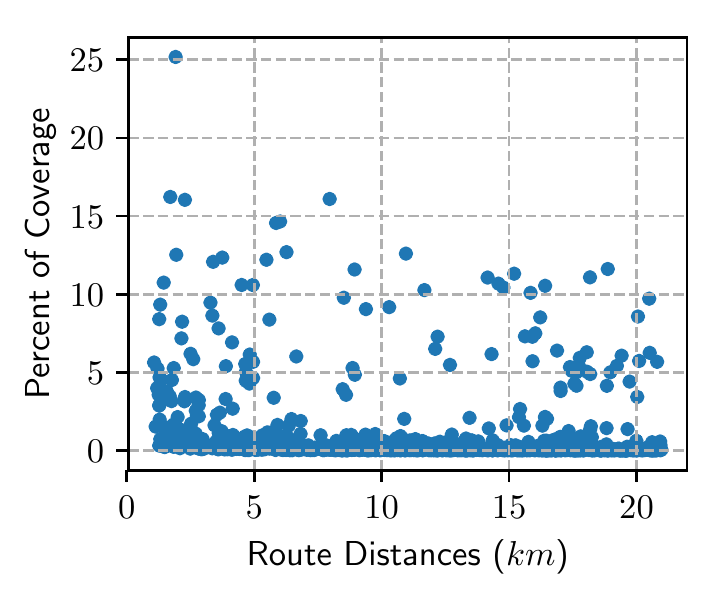}
		\caption{\label{fig:coverage_frankfurt} Frankfurt}
    \end{subfigure}%
    \begin{subfigure}{.25\textwidth}
        \centering
		\includegraphics[width=\textwidth]{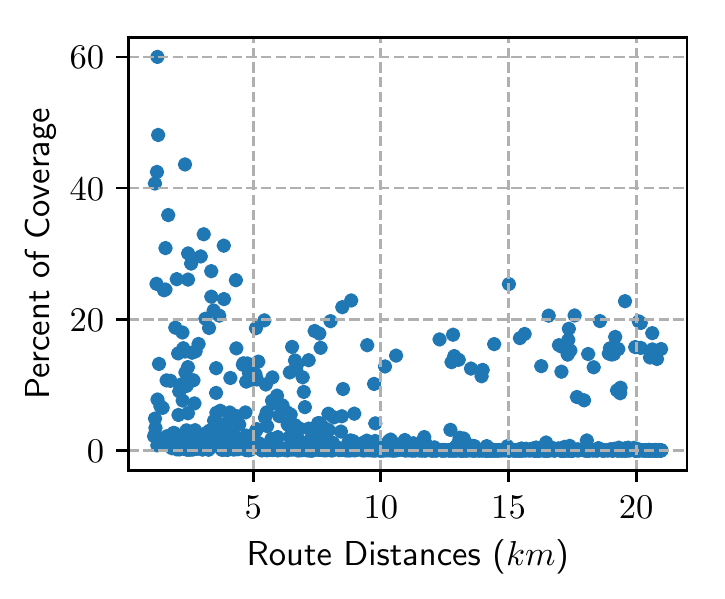}
		\caption{\label{fig:coverage_houston} Houston}
    \end{subfigure}%
    \begin{subfigure}{.25\textwidth}
        \centering
		\includegraphics[width=\textwidth]{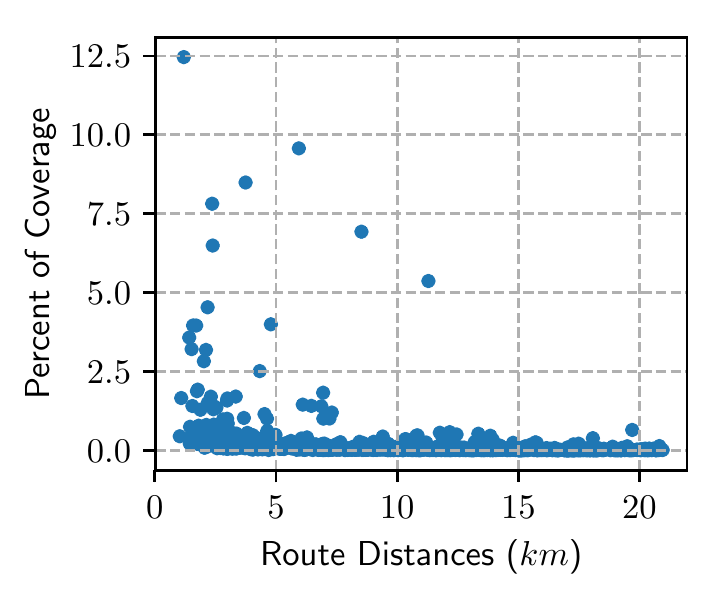}
		\caption{\label{fig:coverage_london} London}
    \end{subfigure}%
    \begin{subfigure}{.25\textwidth}
        \centering
		\includegraphics[width=\textwidth]{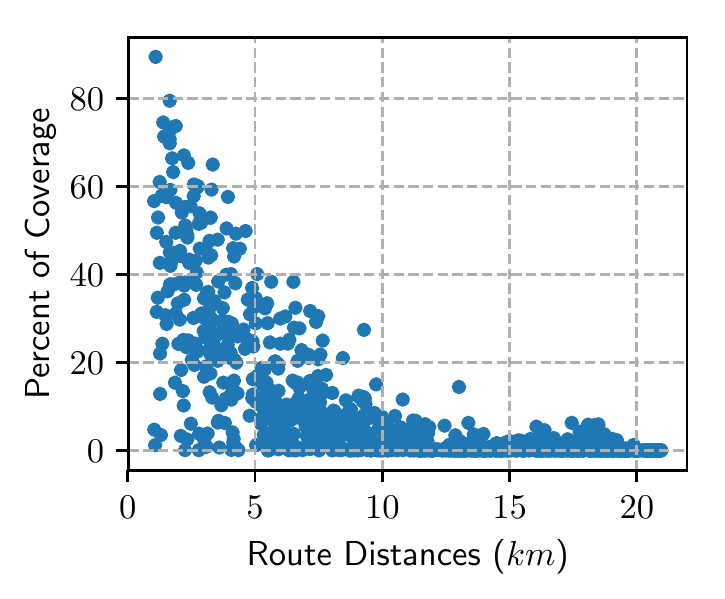}
		\caption{\label{fig:coverage_manhattan} Manhattan}
    \end{subfigure}
    \begin{subfigure}{.25\textwidth}
        \centering
		\includegraphics[width=\textwidth]{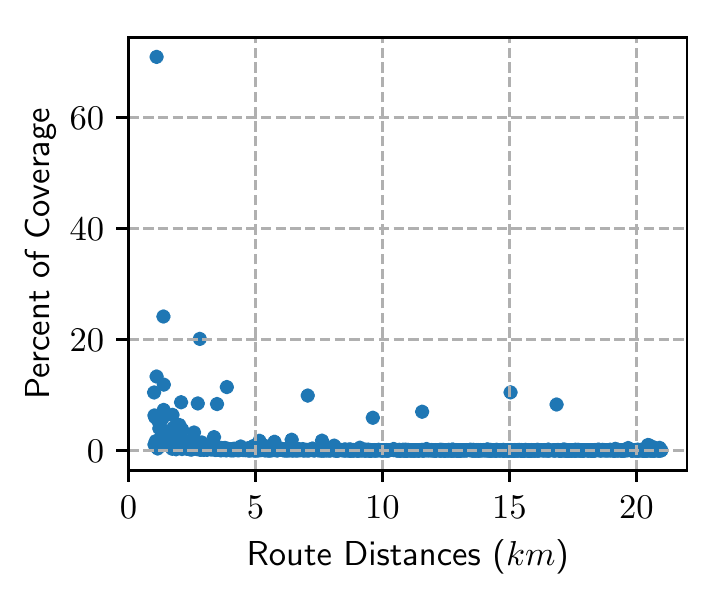}
		\caption{\label{fig:coverage_paris} Paris}
    \end{subfigure}%
    \begin{subfigure}{.25\textwidth}
        \centering
		\includegraphics[width=\textwidth]{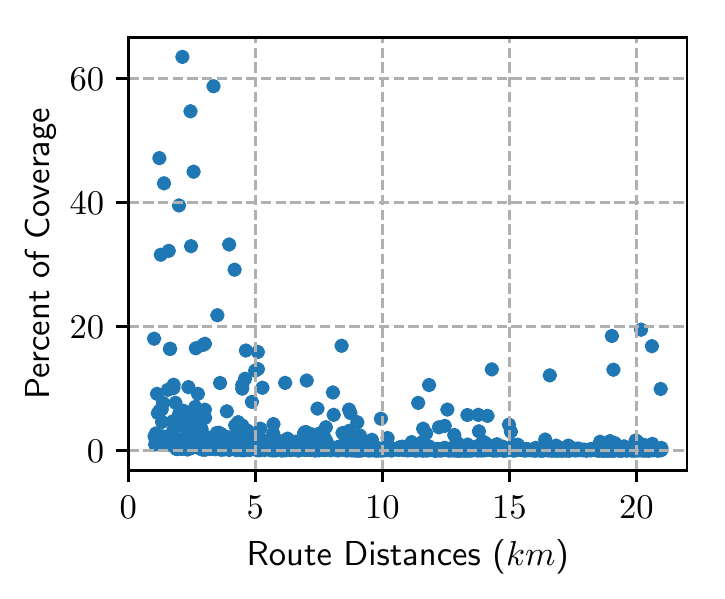}
		\caption{\label{fig:coverage_sf} San Francisco}
    \end{subfigure}
	\caption{Coverage Area of the Attacker: In cities like New York and Chicago, an attacker can cover more than 60\% of the target land area without being detected.}
	\label{fig:coverage}
\end{figure*}

The results are shown in Figure~\ref{fig:coverage}. It can be observed
that cities with more regular grid-like patterns such as Chicago and
Manhattan, New York City are more vulnerable to attacks. It is
possible for an attacker to cover more than 60\% of the target land
area without being detected. However, more irregular cities like
London, Frankfurt and Atlanta offer more resistance. It is important
to note that it is still possible to reach 20\% of the target
geographic region even in these most limiting cases. The percent of
coverage reduces as route or trip distances increases because as trip
length increases so does the probability of the presence of an unique
road segment, but also because the area of interest grows
quadratically in the distance between source and destination.  For
instance, for a distance of $20 km$, the area of interest is $400
km^2$ and the coverage is $40 km^2$ which is still significant.  Also,
note that the above calculations present a lower-bound on the total
coverage area $A_C$. This is because errors in distance calculation
from the accelerometer allows the attacker to cover much larger
distances. For example, in a number of escape routes computed in our
evaluation, up to $82$\% of final escape destinations were located
even beyond the area of interest used for evaluation, with a mean of
$\approx46$\%.


\section{Countermeasures}

The above evaluations demonstrate significant threat of spoofing in urban road networks even when both GPS and the inertial sensors are used together for the localization and tracking. In this section, we present some approaches to mitigate spoofing attacks, specifically in road navigation and tracking applications. 

\subsection{Deploying Accurate Accelerometer and Gyroscope Sensors}
\label{sec:accurate_sensors_impact}

An obvious approach to mitigating the threat would be to use high
quality sensors. To measure the impact of sensor noise on the
potential of spoofing, we re-ran the simulations on the cities using
lower thresholds for the sensor noise. For this evaluation, we set the
thresholds using the $25^{th}$ percentile of the error distributions
(c.f., \Cref{fig:sensors_accuracy}). The following thresholds were set
for the escape paths generator algorithm: $T_\theta = 1.4\degree$,
$T_\vartheta=0.2\degree$, $T_{d1} = 0.6$ and $T_{d2} =
1.6$. \Cref{fig:countermeasure} shows the results of the simulations
for Chicago and San Francisco. Recall that both cities demonstrated
high potential of spoofing for many paths. Using the above thresholds,
we see a significant reduction in the percentage of routes that allow more than $5\unit{km}$ of displacement. However, there
are several limitations with this approach. First, the sensors
satisfying the above parameters are equivalent to aviation and
military-grade sensors which are bulky and expensive (several
thousands of dollars) to deploy. Furthermore, they consume significant
amount of power ($\gtrsim 5\unit{watts}$) making it unsuitable for use
in majority of tracking applications. Moreover, the attacker can still
induce noise in the sensors by driving recklessly such as consistently
switching lanes and accelerating / decelerating.


%

\subsection{Secure Navigation Path Selection} 

\begin{figure}[t]
	\centering
	\begin{subfigure}{.49\linewidth}
    	\includegraphics[width=\textwidth]{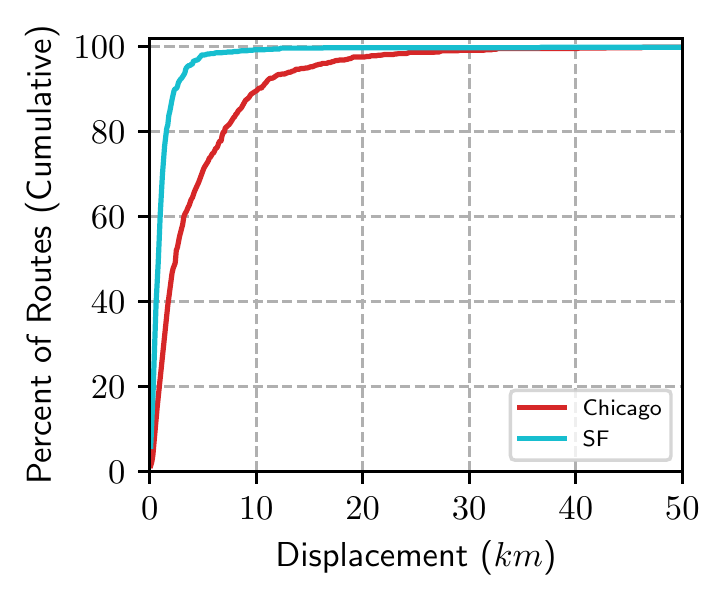}
		\caption{\label{fig:secure_mitigation} Secure Navigation Path Selection}
    \end{subfigure}
    \begin{subfigure}{.49\linewidth}
        \centering
		\includegraphics[width=\textwidth]{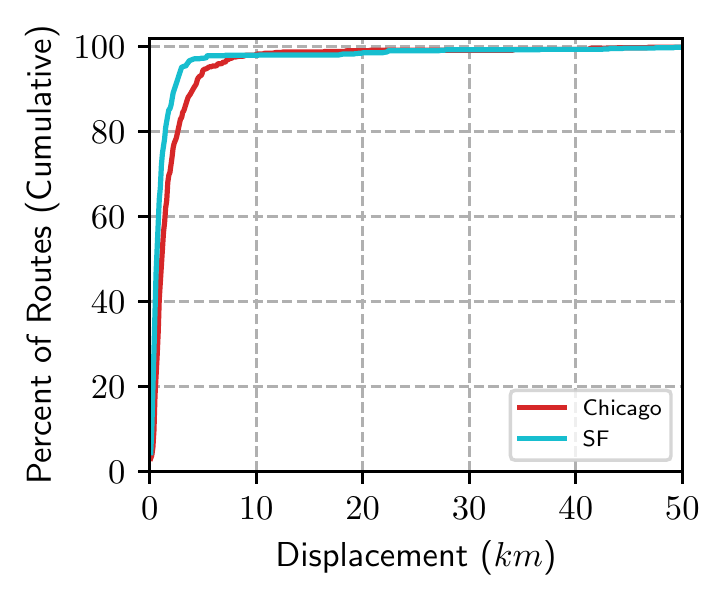}
		\caption{\label{fig:sensor_mitigation} Military- or Aviation-grade sensors}
    \end{subfigure}
    \caption{Preliminary results of countermeasure: We see that both using higher accuracy sensors (expensive, bulky, high power) and our secure navigation path selection (easy to deploy) significantly reduces the impact of the attack. }
	\label{fig:countermeasure}
\end{figure}

Recall that the attack algorithm searches for navigation routes between the assigned start and end points containing attributes with the high probability of occurrence in other parts of the road network i.e., other sections of the graph (c.f., \Cref{sec:algo1}). The final path score was calculated using \Cref{eq:path_score}. The idea behind generating paths more resilient to spoofing is to simply negate this path score, i.e., $score = -(\prod_{i=1}^{M} P(\vartheta(s_i), \theta(\chi_i)))$. This has the effect of assigning the highest score to a path containing road curvature and turn angles with low probability of occurrence. These paths are less favorable for spoofing because the curvatures or turn angles in the path are more unique and, therefore, less likely in other sections. The algorithm uses the same inputs as the previous algorithm but sets the count of output paths $N_P$ as $1$, i.e., it outputs the most secure path it finds for the given source and destination. In other words, the application or service provider (\eg logistics company) can assign ``secure navigation routes'' that are hard to fake because of unique road characteristics. \Cref{fig:countermeasure} shows the results of a preliminary evaluations for Chicago and San Francisco. Comparing with the original simulations, we again see that the attacker is significantly limited in the amount of routes available to him for reaching alternate destinations. 

The key advantage of our secure navigation path algorithm is that there is no changes needed to the existing GPS/INS hardware tracking required. The company can simply choose the ``secure path'' to travel instead of deploying new sensors for every tracking device. Furthermore, even if there exists some potential for spoofing in the best possible secure path, the escape routes can be known well in advance and appropriate countermeasure be taken to prevent it.


\section{Related Work}
In this section we discuss relevant related work beginning with prior works that have demonstrated various attacks on GPS. In 2001, the Volpe report~\cite{volpe2001vulnerability} first identified malicious interference with the civilian GPS signal as a serious problem. Following this several researchers have demonstrated the insecurity of GPS-based navigation by diverting the course of a yacht~\cite{yacht_spoofing}, forcing drones~\cite{humphreys2012statement} to land in a hostile area and taken over navigation systems of transportation trucks~\cite{Warner2003} using spoofed GPS signals. More recently, researchers demonstrated a GPS signal generator that can be built for less than \$300~\cite{lowcost_gps}. Today, there exist public software repositories~\cite{gps-sdr-sim} as well as commercial GPS simulators~\cite{labsat,spectracom} that generate GPS signals for any chosen trajectory or navigation route. More advanced attacks were demonstrated in~\cite{NighswanderCCS2012,Tippenhauer2011} in which the attackers \textit{takeover} a target receiver that is already locked onto (\ie continuously receiving navigation messages) authentic satellite signals without the receiver noticing any disruption or loss of navigation data. It was also shown that a variety of commercial GPS receivers were vulnerable and in some cases even caused permanent damage to the receivers. 

A number of countermeasures have been proposed against GPS spoofing attacks. Several works~\cite{kuhn2005asymmetric,lo2010authenticating,wesson2012practical} proposed solutions that are cryptographic in nature and therefore require modifications to the GPS infrastructure. Many non-cryptographic countermeasures  rely on detecting anomalies in certain physical characteristics of the signal such as received satellite signal strength~\cite{Warner2003}, ambient noise floor levels, automatic gain control values~\cite{Akos2012} and other data that are readily available as receiver observables on modern GPS receivers. Some other countermeasures~\cite{Psiaki2011,phelts2001multicorrelator,Wesson2011} leveraged the signal's spatial characteristics such as the received GPS signal's direction or angle of arrival. Some proposed and analyzed the use of multiple synchronized GPS receivers~\cite{Tippenhauer2011,swaszek2013multireceiver,swaszek2014multireceiver} to detect spoofing. They show that spoofing a set of synchronized GPS receivers, with known relative distances or geometrical constellation restricts the number of locations from
where an attacker can transmit the spoofing signals. Some other works~\cite{ranganathan2016spree} leveraged the difficulty of completely annihilating legitimate signals from the environment. Cross-validation of the position estimates against alternate navigation systems such as Galileo~\cite{hofmann2007gnss} were also proposed. All the above countermeasures require modifications to the GPS infrastructure or receiver. The multi receiver solutions require the receivers to be at least 5--6 m away from each other making them unsuitable for road navigation applications. 

In the context of road navigation and tracking, using data from inertial sensors~\cite{titterton2004strapdown,farrell1999global,wendel2006integrated} alongside GPS is emerging as a popular choice for tracking and navigation in applications where spoofing and jamming are considered a threat. The absence of any communication between the inertial sensors and the external world for estimating the location makes it robust to signal spoofing and jamming attacks. Many works~\cite{khanafseh2014gps,white1998detection,lee2015gps,dehnie2014methods,ebner1997integrated,serrano2011receiver} analyze and show that inertial sensors are promising for detection and mitigation of GPS spoofing attacks. Many commercial-off-the-shelf GPS/INS products~\cite{kvh_imu,vectorNAV_imu,honeywell_IMU,navtech_gpsimu} are available and used in many civilian and military applications. Recently, analog attacks have also been demonstrated on inertial sensors. For example, WALNUT~\cite{trippel2017walnut} shows how analog acoustic injection attacks can affect the digital integrity of a capacitive MEMS accelerometer. Son et al.~\cite{son2015rocking} showed that acoustic interference on MEMS gyroscopes in drones can cause them to crash. In~\cite{shoukry2013non}, Shoukry et al. demonstrate how to deliver fake readings to a anti-lock braking systems (ABS) via the magnetic wheel speed sensors using electro magnetic interference in an automotive setting. In this paper, we   show that magnetometers are vulnerable to electromagnetic interference attacks and an attacker can precisely control its output.

Given the emergence of GPS/INS solutions, we believe our work emphasizes some fundamental security limitations of GPS/INS for road navigation and tracking applications.

\section{Conclusion}

In this paper, we evaluated the security guarantees of GPS/INS based
tracking and navigation for road transportation systems. To this
extent, we designed a suite of algorithms that enable an attacker to
derive escape routes and plausible destinations to reach without
raising alarms even with the INS-aided GPS tracking and navigation
system. We implemented and evaluated the impact of the attack using
both real-world and simulated driving traces in more than 10 cities
located around the world and showed that is possible for an attacker
to evade detection and reach locations that are as far as 30 km away
from the true destination and, in some cases, cover more than 60\% of
the target geographic region. Finally, we proposed countermeasures
that do not require any hardware modifications and yet can severely
limit the attacker's ability to cheat.

\bibliographystyle{IEEEtran}
\bibliography{insgpsmap}

\begin{thebibliography}{10}
\providecommand{\url}[1]{#1}
\csname url@samestyle\endcsname
\providecommand{\newblock}{\relax}
\providecommand{\bibinfo}[2]{#2}
\providecommand{\BIBentrySTDinterwordspacing}{\spaceskip=0pt\relax}
\providecommand{\BIBentryALTinterwordstretchfactor}{4}
\providecommand{\BIBentryALTinterwordspacing}{\spaceskip=\fontdimen2\font plus
\BIBentryALTinterwordstretchfactor\fontdimen3\font minus
  \fontdimen4\font\relax}
\providecommand{\BIBforeignlanguage}[2]{{%
\expandafter\ifx\csname l@#1\endcsname\relax
\typeout{** WARNING: IEEEtran.bst: No hyphenation pattern has been}%
\typeout{** loaded for the language `#1'. Using the pattern for}%
\typeout{** the default language instead.}%
\else
\language=\csname l@#1\endcsname
\fi
#2}}
\providecommand{\BIBdecl}{\relax}
\BIBdecl

\bibitem{coffee2003vehicle}
J.~R. Coffee, R.~W. Rudow, R.~F. Allen, M.~Billings, D.~A. Dye, M.~L. Kirchner,
  R.~W. Lewis, K.~M. Marvin, R.~D. Sleeper, W.~A. Tekniepe \emph{et~al.},
  ``Vehicle tracking, communication and fleet management system,'' Aug.~26
  2003, uS Patent 6,611,755.

\bibitem{novik2002system}
Y.~A. Novik, ``System and method for fleet tracking,'' Jan.~15 2002, uS Patent
  6,339,745.

\bibitem{verizon-fleet-connect}
``{Verizon Connect Fleet Management System},''
  \url{https://www.verizonconnect.com/solutions/gps-fleet-tracking-software/}.

\bibitem{ma-electronicmonitoring}
``{Massachusetts Probation Service's Electronic monitoring program },''
  \url{https://www.mass.gov/service-details/electronic-monitoring-program}.

\bibitem{geosatis}
``{Geo-Satis Electronic Monitoring Solution},'' \url{https://geo-satis.com/}.

\bibitem{masstransit-usa}
``{US Department of Transportation: In-vehicle Performance Monitoring and
  Feedback},''
  \url{https://www.transportation.gov/mission/health/In-vehicle-Performance-Monitoring-and-Feedback}.

\bibitem{mintsis2004applications}
G.~Mintsis, S.~Basbas, P.~Papaioannou, C.~Taxiltaris, and I.~Tziavos,
  ``Applications of gps technology in the land transportation system,''
  \emph{European journal of operational Research}, vol. 152, no.~2, pp.
  399--409, 2004.

\bibitem{civitas_eu}
``{Developing GPS monitoring for the public transport fleet},''
  \url{http://civitas.eu/measure/developing-gps-monitoring-public-transport-fleet}.

\bibitem{misra2006global}
P.~Misra and P.~Enge, \emph{Global Positioning System: Signals, Measurements
  and Performance Second Edition}.\hskip 1em plus 0.5em minus 0.4em\relax
  Lincoln, MA: Ganga-Jamuna Press, 2006.

\bibitem{gsa2017market}
G.~GSA, ``Market report issue 3,'' \url{https://www.gsa.europa.eu/}.

\bibitem{yacht_spoofing}
``{UT Austin Researchers Successfully Spoof an \$80 million Yacht at Sea},''
  http://news.utexas.edu/2013/07/29/ut-austin-researchers-successfully-spoof-an-80-million-yacht-at-sea.

\bibitem{humphreys2012statement}
T.~Humphreys, ``Statement on the vulnerability of civil unmanned aerial
  vehicles and other systems to civil gps spoofing,'' \emph{University of Texas
  at Austin (July 18, 2012)}, 2012.

\bibitem{zeng2017practical}
K.~C. Zeng, Y.~Shu, S.~Liu, Y.~Dou, and Y.~Yang, ``A practical gps location
  spoofing attack in road navigation scenario,'' in \emph{Proceedings of the
  18th International Workshop on Mobile Computing Systems and
  Applications}.\hskip 1em plus 0.5em minus 0.4em\relax ACM, 2017, pp. 85--90.

\bibitem{gizmodo_gpsjamming}
``{Jamming GPS Signals Is Illegal, Dangerous, Cheap, and Easy},''
  \url{https://gizmodo.com/jamming-gps-signals-is-illegal-dangerous-cheap-and-e-1796778955}.

\bibitem{cbs_gpsjamming}
``{N.J. Man In A Jam, After Illegal GPS Device Interferes With Newark Liberty
  Operations},''
  https://newyork.cbslocal.com/2013/08/09/n-j-man-in-a-jam-after-illegal-gps-device-interferes-with-newark-liberty-operations/.

\bibitem{humphreys2013detection}
T.~E. Humphreys, ``{Detection strategy for cryptographic GNSS anti-spoofing},''
  \emph{IEEE Transactions on Aerospace and Electronic Systems}, 2013.

\bibitem{kuhn2005asymmetric}
M.~G. Kuhn, ``An asymmetric security mechanism for navigation signals,'' in
  \emph{Information Hiding}, 2005.

\bibitem{lo2010authenticating}
S.~C. Lo and P.~K. Enge, ``Authenticating aviation augmentation system
  broadcasts,'' 2010.

\bibitem{wesson2012practical}
K.~Wesson, M.~Rothlisberger, and T.~Humphreys, ``{Practical cryptographic civil
  GPS signal authentication},'' \emph{Journal of Navigation}, 2012.

\bibitem{Akos2012}
D.~M. Akos, ``{Who's afraid of the spoofer? GPS/GNSS spoofing detection via
  automatic gain control (AGC)},'' \emph{Navigation}, 2012.

\bibitem{ranganathan2016spree}
A.~Ranganathan, H.~{\'O}lafsd{\'o}ttir, and S.~Capkun, ``Spree: A spoofing
  resistant gps receiver,'' in \emph{Proceedings of the 22nd Annual
  International Conference on Mobile Computing and Networking}.\hskip 1em plus
  0.5em minus 0.4em\relax ACM, 2016, pp. 348--360.

\bibitem{psiaki2013gnss}
M.~L. Psiaki, S.~P. Powell, and B.~W. O'Hanlon, ``{GNSS spoofing detection
  using high-frequency antenna motion and carrier-phase data},'' in
  \emph{Proceedings of the ION GNSS+ Meeting}, 2013.

\bibitem{Wesson2011}
K.~Wesson, D.~Shepard, J.~Bhatti, and T.~E. Humphreys, ``{An evaluation of the
  vestigial signal defense for civil GPS anti-spoofing},'' in \emph{Proceedings
  of the ION GNSS Meeting}, 2011.

\bibitem{Warner2003}
J.~S. Warner and R.~G. Johnston, ``{GPS spoofing countermeasures},''
  \emph{Homeland Security Journal}, 2003.

\bibitem{broumandan2012gnss}
A.~Broumandan, A.~Jafarnia-Jahromi, V.~Dehghanian, J.~Nielsen, and
  G.~Lachapelle, ``{GNSS spoofing detection in handheld receivers based on
  signal spatial correlation},'' in \emph{Proceedings of the IEEE Position
  Location and Navigation Symposium (PLANS)}, 2012.

\bibitem{Jafarnia-Jahromi2012}
A.~Jafarnia-Jahromi, A.~Broumandan, J.~Nielsen, and G.~Lachapelle, ``{GPS
  vulnerability to spoofing threats and a review of antispoofing techniques},''
  \emph{International Journal of Navigation and Observation}, 2012.

\bibitem{zandbergen2009accuracy}
P.~A. Zandbergen, ``Accuracy of iphone locations: A comparison of assisted gps,
  wifi and cellular positioning,'' \emph{Transactions in GIS}, vol.~13, pp.
  5--25, 2009.

\bibitem{tippenhauer2009attacks}
N.~O. Tippenhauer, K.~B. Rasmussen, C.~P{\"o}pper, and S.~{\v{C}}apkun,
  ``Attacks on public wlan-based positioning systems,'' in \emph{Proceedings of
  the 7th international conference on Mobile systems, applications, and
  services}.\hskip 1em plus 0.5em minus 0.4em\relax ACM, 2009, pp. 29--40.

\bibitem{titterton2004strapdown}
D.~Titterton, J.~Weston \emph{et~al.}, \emph{{Strapdown Inertial Navigation
  Technology. 2nd Edition}}.\hskip 1em plus 0.5em minus 0.4em\relax IET, 2004.

\bibitem{farrell1999global}
J.~Farrell and M.~Barth, \emph{{The Global Positioning System and inertial
  navigation}}.\hskip 1em plus 0.5em minus 0.4em\relax McGraw-Hill New York,
  1999.

\bibitem{wendel2006integrated}
J.~Wendel, O.~Meister, C.~Schlaile, and G.~F. Trommer, ``{An integrated
  GPS/MEMS-IMU navigation system for an autonomous helicopter},''
  \emph{Aerospace Science and Technology}, 2006.

\bibitem{kvh_imu}
``{KVH Systems - Using Inertial Systems to Overcome GPS Spoofing},''
  \url{https://www.kvhmobileworld.kvh.com/}.

\bibitem{vectorNAV_imu}
``{VectorNAV - Embedded Navigation Solutions},''
  \url{https://www.vectornav.com/products}.

\bibitem{honeywell_IMU}
``{Honeywell Aerospace - Embedded GPS/INS},''
  \url{https://aerospace.honeywell.com/en/products/navigation-and-sensors/embedded-gps-or-ins}.

\bibitem{navtech_gpsimu}
``{Navtech GPS soplutions},''
  \url{https://www.navtechgps.com/oxts_xoem_inseries/}.

\bibitem{khanafseh2014gps}
S.~Khanafseh, N.~Roshan, S.~Langel, F.-C. Chan, M.~Joerger, and B.~Pervan,
  ``Gps spoofing detection using raim with ins coupling,'' in \emph{Proceedings
  of the Position, Location and Navigation Symposium—PLANS}, vol. 2014, 2014.

\bibitem{white1998detection}
N.~A. White, P.~S. Maybeck, and S.~L. DeVilbiss, ``Detection of
  interference/jamming and spoofing in a dgps-aided inertial system,''
  \emph{IEEE Transactions on Aerospace and Electronic Systems}, vol.~34, no.~4,
  pp. 1208--1217, 1998.

\bibitem{lee2015gps}
J.-H. Lee, K.-C. Kwon, D.-S. An, and D.-S. Shim, ``Gps spoofing detection using
  accelerometers and performance analysis with probability of detection,''
  \emph{International Journal of Control, Automation and Systems}, vol.~13,
  no.~4, pp. 951--959, 2015.

\bibitem{dehnie2014methods}
S.~Dehnie and R.~Ghanadan, ``Methods and systems for detecting gps spoofing
  attacks,'' Dec.~30 2014, uS Patent 8,922,427.

\bibitem{ebner1997integrated}
R.~E. Ebner and R.~A. Brown, ``Integrated gps/inertial navigation apparatus
  providing improved heading estimates,'' Aug.~12 1997, uS Patent 5,657,025.

\bibitem{serrano2011receiver}
L.~M. P.~A. Serrano, C.~S. Dixon, and M.~J. Perren, ``Receiver and method for
  authenticating satellite signals,'' Jul.~28 2011, uS Patent App. 12/780,337.

\bibitem{ettus_research}
``Ettus research llc,'' \url{http://www.ettus.com/}.

\bibitem{lowcost_gps}
``{Hacking A Phone's GPS May Have Just Got Easier},''
  \url{http://www.forbes.com/sites/parmyolson/2015/08/07/gps-spoofing-hackers-defcon/}.

\bibitem{gps-sdr-sim}
``{Opensource software-defined GPS signal simulator},''
  \url{https://github.com/osqzss/gps-sdr-sim}.

\bibitem{rebeiz2004rf}
G.~M. Rebeiz, \emph{RF MEMS: theory, design, and technology}.\hskip 1em plus
  0.5em minus 0.4em\relax John Wiley \& Sons, 2004.

\bibitem{osm}
{OpenStreetMap}, ``{OpenStreetMap Project},'' https://www.openstreetmap.org/.

\bibitem{narain_snp16}
S.~Narain, T.~D. Vo-Huu, K.~Block, and G.~Noubir, ``Inferring user routes and
  locations using zero-permission mobile sensors,'' in \emph{2016 IEEE
  Symposium on Security and Privacy (SP)}, May 2016.

\bibitem{volpe2001vulnerability}
J.~A. Volpe, ``Vulnerability assessment of the transportation infrastructure
  relying on the global positioning system,'' \emph{http://www. navcen. uscg.
  gov/}, 2001.

\bibitem{labsat}
``{LabSat GPS Simulator},'' \url{http://www.labsat.co.uk/}.

\bibitem{spectracom}
``{GSG-xx Series Multi-channel advanced GNSS simulator},''
  \url{http://www.spectracomcorp.com/}.

\bibitem{NighswanderCCS2012}
T.~Nighswander, B.~M. Ledvina, J.~Diamond, R.~Brumley, and D.~Brumley, ``{GPS
  software attacks},'' in \emph{Proceedings of the ACM Conference on Computer
  and Communications Security}, 2012.

\bibitem{Tippenhauer2011}
N.~O. Tippenhauer, C.~P{\"o}pper, K.~B. Rasmussen, and S.~Capkun, ``{On the
  requirements for successful GPS spoofing attacks},'' in \emph{Proceedings of
  the 18th ACM Conference on Computer and communications security}, 2011.

\bibitem{Psiaki2011}
M.~L. Psiaki, B.~W. O'Hanlon, J.~A. Bhatti, D.~P. Shepard, and T.~E. Humphreys,
  ``{Civilian GPS spoofing detection based on dual-receiver correlation of
  military signals},'' \emph{Institute of Navigation GNSS (ION GNSS)}, 2011.

\bibitem{phelts2001multicorrelator}
R.~E. Phelts, ``Multicorrelator techniques for robust mitigation of threats to
  gps signal quality,'' Ph.D. dissertation, Stanford University, 2001.

\bibitem{swaszek2013multireceiver}
H.~R. J. K. M. V. J. G.~W. Swaszek, Peter~F., ``Analysis of a simple,
  multi-receiver gps spoof detector,'' in \emph{Proceedings of the 2013
  International Technical Meeting of The Institute of Navigation}.\hskip 1em
  plus 0.5em minus 0.4em\relax ION, 2013.

\bibitem{swaszek2014multireceiver}
H.~R. Swaszek, P.F., ``A multiple cots receiver gnss spoof detector --
  extensions,'' in \emph{Proceedings of the 2014 International Technical
  Meeting of The Institute of Navigation}.\hskip 1em plus 0.5em minus
  0.4em\relax ION, 2014.

\bibitem{hofmann2007gnss}
B.~Hofmann-Wellenhof, H.~Lichtenegger, and E.~Wasle, \emph{{GNSS--global
  navigation satellite systems: GPS, GLONASS, Galileo, and more}}.\hskip 1em
  plus 0.5em minus 0.4em\relax Springer Science \& Business Media, 2007.

\bibitem{trippel2017walnut}
T.~Trippel, O.~Weisse, W.~Xu, P.~Honeyman, and K.~Fu, ``Walnut: Waging doubt on
  the integrity of mems accelerometers with acoustic injection attacks,'' in
  \emph{Security and Privacy (EuroS\&P), 2017 IEEE European Symposium
  on}.\hskip 1em plus 0.5em minus 0.4em\relax IEEE, 2017, pp. 3--18.

\bibitem{son2015rocking}
Y.~M. Son, H.~C. Shin, D.~K. Kim, Y.~S. Park, J.~H. Noh, K.~B. Choi, J.~W.
  Choi, and Y.~D. Kim, ``Rocking drones with intentional sound noise on
  gyroscopic sensors,'' in \emph{24th USENIX Security symposium}.\hskip 1em
  plus 0.5em minus 0.4em\relax USENIX Association, 2015.

\bibitem{shoukry2013non}
Y.~Shoukry, P.~Martin, P.~Tabuada, and M.~Srivastava, ``Non-invasive spoofing
  attacks for anti-lock braking systems,'' in \emph{International Workshop on
  Cryptographic Hardware and Embedded Systems}.\hskip 1em plus 0.5em minus
  0.4em\relax Springer, 2013, pp. 55--72.

\end{thebibliography}

\end{document}